\documentclass[twocolumn, tighten, times, twocolappendix]{aastex}
\usepackage{amsmath}
\usepackage{microtype}
\usepackage{siunitx}

\graphicspath{{figures/}}

\newcommand{\mej}{\ensuremath{M_\mathrm{ej}}}
\newcommand{\kh}{Kelvin--Helmholtz} %
\newcommand{\crosssection}{cross section}
\newcommand{\stefanboltzmann}{Stefan--Boltzmann} %

\newcommand{\rjup}{\ensuremath{R_\mathrm{J}}}

\newcommand{\lsun}{\ensuremath{L_\odot}}

\newcommand{\mplanet}{\ensuremath{M_\mathrm{p}}}
\newcommand{\rplanet}{\ensuremath{R_\mathrm{p}}}
\newcommand{\menv}{\ensuremath{M_\mathrm{env}}}
\newcommand{\renv}{\ensuremath{R_\mathrm{env}}}

\newcommand{\theat}{\ensuremath{\tau_\mathrm{heat}}}
\newcommand{\eorb}{\ensuremath{E_\mathrm{orb}}}
\newcommand{\porb}{\ensuremath{P_\mathrm{orb}}}
\newcommand{\rstar}{\ensuremath{R_\star}}
\newcommand{\rsun}{\ensuremath{R_\odot}}
\newcommand{\mstar}{\ensuremath{M_\star}}

\newcommand{\lstar}{\ensuremath{L_\star}}

\newcommand{\edottide}{\ensuremath{\dot{E}_\mathrm{tide}}}
\newcommand{\cs}{\ensuremath{c_\mathrm{s}}}

\newcommand{\ttide}{\ensuremath{\tau_\mathrm{tide}}}

\newcommand{\tdrag}{\ensuremath{\tau_\mathrm{drag}}}
\newcommand{\teff}{\ensuremath{T_\mathrm{eff}}}
\newcommand{\period}{\ensuremath{P_\mathrm{orb}}}
\newcommand{\vorb}{\ensuremath{v_\mathrm{orb}}}
\newcommand{\vesc}{\ensuremath{v_\mathrm{esc}}}
\newcommand{\qstar}{\ensuremath{Q_\star^\prime}}

\newcommand{\qpenev}{\ensuremath{Q_{\star,\mathrm{Penev}}^\prime}}
\newcommand{\event}{ZTF SLRN-2020}

\newcommand{\epsrho}{\ensuremath{\varepsilon_\rho}}
\newcommand{\hrho}{\ensuremath{H_\rho}}
\newcommand{\ieint}{\ensuremath{I_\mathrm{e}}}
\newcommand{\imint}{\ensuremath{I_\mathrm{m}}}

\newcommand{\lp}{\ensuremath{\left(}} %
\newcommand{\rp}{\ensuremath{\right)}} %

\DeclareSIUnit{\year}{yr}
\DeclareSIUnit{\day}{d}
\DeclareSIUnit{\dex}{dex}
\DeclareSIUnit{\gauss}{G}
\DeclareSIUnit{\erg}{erg}
\DeclareSIUnit{\parsec}{pc}
\DeclareSIUnit{\solarmass}{M_\odot}
\DeclareSIUnit{\earthmass}{M_\oplus}
\DeclareSIUnit{\earthradius}{R_\oplus}
\DeclareSIUnit{\solarluminosity}{L_\odot}
\DeclareSIUnit{\solarradius}{R_\odot}
\DeclareSIUnit{\jupitermass}{M_{J}}
\DeclareSIUnit{\jupiterradius}{R_{J}}

\begin{document}
\title{Evolution of the \event{} star-planet merger}
\shorttitle{Evolution of the \event{} star-planet merger}
\shortauthors{Yarza et al.}

\author[0000-0003-0381-1039]{Ricardo~Yarza} %
\email{ryarza@ucsc.edu}
\altaffiliation{NASA FINESST Fellow}
\affiliation{Department of Astronomy and Astrophysics, University of California, Santa Cruz, CA 95064, USA}

\author[0000-0002-1417-8024]{Morgan~MacLeod} %
\email{morgan.macleod@cfa.harvard.edu}
\affiliation{Center for Astrophysics, Harvard \& Smithsonian, 60 Garden Street, MS-16, Cambridge, MA 02138, USA}
\affiliation{Institute for Theory and Computation, Harvard \& Smithsonian, 60 Garden Street, MS-51, Cambridge, MA 02138, USA}

\author[0000-0002-2697-3893]{Benjamin~Idini} %
\email{bidini@ucsc.edu}
\affiliation{Department of Astronomy and Astrophysics, University of California, Santa Cruz, CA 95064, USA}

\author[0000-0001-5061-0462]{Ruth~Murray-Clay} %
\email{rmc@ucsc.edu}
\affiliation{Department of Astronomy and Astrophysics, University of California, Santa Cruz, CA 95064, USA}

\author[0000-0003-2558-3102]{Enrico~Ramirez-Ruiz} %
\email{enrico@ucolick.org}
\affiliation{Department of Astronomy and Astrophysics, University of California, Santa Cruz, CA 95064, USA}

\begin{abstract}
We model the optical and infrared transient \event{}, previously associated with a star-planet merger. We consider the scenario in which orbital decay via tidal dissipation led to the merger, and find that tidal heating within the star was likely unobservable in the archival image of the system taken \qty{12}{\year} before the merger. The observed dust formation months before the merger is consistent with a planet of mass \( \mplanet \gtrsim \qty{5}{\jupitermass} \) ejecting material as it skims the stellar surface. This interaction gradually intensifies, leading to significant mass ejection on a dynamical timescale (\( \approx \) hours) as the planet plunges into the stellar interior. Part of the recombination transient associated with this dynamical mass ejection might be inaccessible to the optical observations because its duration (\( \approx \) hours) is comparable to the cadence. Correspondingly, the observed duration of the transient \(\approx\qty{100}{\day}\) is inconsistent with a single episode of dynamical mass ejection. Instead, the transient could be powered by the recombination of \( \qty{3.4e-5}{\solarmass} \) of hydrogen in an outflow, or the contraction of an inflated envelope of mass \( \approx \qty{e-6}{\solarmass} \) that formed during the merger. The observed ejecta mass \qty{320}{\day} after the peak of the optical transient is \( \approx \qty{1.3e-4}{\solarmass} \), consistent with the idea that a fraction of the ejecta might be unobservable in the light curve. Energetically, this post-merger ejecta mass suggests a planet at least as massive as Jupiter. Our results suggest that \event{} was the result of a merger between a star close to the main sequence and a planet with mass at least several times that of Jupiter.
\end{abstract}

\section{Introduction}
The observed orbital configurations of planetary systems imply that a large fraction of the known exoplanets will merge with their host stars at some point during their evolution \citep{Rasio1996,Villaver2007,Villaver2009,Jackson2009,Levrard2009,Matsumura2010,Nordhaus2010,Carlberg2011a,Kunitomo2011,Metzger2012,Mustill2012,Nordhaus2013,Schlaufman2013,Villaver2014,Matsakos2015,Damiani2016,Veras2016,Rapoport2021,Kane2023,Mustill2024}. The main mechanisms leading to these mergers are tidal dissipation of orbital energy, stellar expansion during the post-main-sequence, and dynamical interactions. Roughly 0.5\% of sunlike stars host a hot Jupiter \citep{Howard2012}. The occurrence rate of hot Jupiters appears to decrease with stellar age along the main sequence, suggesting that tidal dissipation is efficient enough to lead to mergers on timescales comparable to the main sequence lifetime \citep{Hamer2019,Miyazaki2023}. On average, each sunlike star has \(\approx 1 \) planet with radius between \( \qty{1}{\earthradius} \) and \( \qty{20}{\earthradius} \) and period \(<\qty{400}{\day} \) \citep{Zhu2021}; at such orbital periods, tides and post-main-sequence expansion often lead to mergers. Estimates for the merger rate in the Galaxy range from one every few years to a few per year \citep{Metzger2012,MacLeod2018,Popkov2019,De2023}.

A star-planet merger potentially produces a range of observable effects. The angular momentum of the orbit of the planet is transferred to the star, increasing its spin rate \citep{Siess1999,Siess1999a,Livio2002,Massarotti2008,Carlberg2009,Carlberg2011a,Carlberg2012,Carlberg2013,Zhang2014,Privitera2016a,Qureshi2018,Oetjens2020,Stephan2020,Cabezon2023,Guo2023,Lau2025}. The deposition of planetary material in the star temporarily changes the stellar atmospheric abundances \citep{Alexander1967,Laughlin1997,Sandquist1998,Siess1999,Siess1999a,Gratton2001,Montalban2002,Sandquist2002,Cody2005,Carlberg2012,Carlberg2013,AguileraGomez2016,AguileraGomez2016a,Nagar2020,SoaresFurtado2021,Sevilla2022,Cabezon2023,Xie2023,Lau2025,OConnor2025,Soares2025}, which could explain unusual abundance patterns in a few systems \citep{Israelian2001,Li2008,Carlberg2010,Adamow2012,Spina2015,Tognelli2016,Saffe2017,Oh2018,Church2020,YanaGalarza2021,Miquelarena2024,YanaGalarza2024}. Broadly, \( \approx5\% \) of main-sequence stars have abundance patterns consistent with a previous merger with a planet \citep{Behmard2023,Liu2024}. A star-planet merger could also change the magnetic field of the star \citep{Siess1999a,Privitera2016b}. Overall, the timescales over which these rotational, magnetic, and chemical signatures remain observable depend on mixing and angular momentum transport within the star, two processes that remain poorly understood \citep{Privitera2016a,Sevilla2022}.

Until recently, the observational evidence for star-planet mergers had been exclusively indirect: present-day orbital configurations imply future mergers, while stars with observed anomalous properties suggest past mergers, with no directly observed connection from pre- to post-merger systems.
The indirect evidence for star-planet mergers suffers from the fact that many non-merger processes can produce signatures similar to mergers. For example, stellar chemical anomalies can result from stellar internal processes \citep{Cameron1971,Carlberg2013,Yan2018,Casey2019,AguileraGomez2020,Sayeed2024} or the intrinsic variability in stellar compositions \citep{Behmard2023,Saffe2024,Soliman2025,Sun2025}. These challenges further motivate the search for the transients produced by star-planet mergers in real time.

Multiple studies have modeled the transients produced by star-planet mergers \citep[e.g.,][]{Bear2011a,Metzger2012,Kashi2017,Metzger2017,Yamazaki2017,MacLeod2018,Kashi2019,Matsumoto2022,OConnor2023}. The primary mechanism responsible for optical emission is the recombination of hydrogen in an outflow of increasing strength as the star and planet approach coalescence over the course of \( \approx \) weeks \citep{Metzger2012}. When the planet finally plunges into the stellar interior on a dynamical timescale \( \approx \) hours, a mass ejection event likely produces a bright recombination transient with a duration comparable to a few orbital periods \citep{Metzger2012,Yamazaki2017}. The star radiates energy deposited deeper in the interior over the course of the much longer \kh{} time \citep{Metzger2012,Metzger2017,MacLeod2018,OConnor2023}. Hydrodynamical simulations are consistent with this overall picture \citep{Sandquist2002,Staff2016,Kramer2020,Cabezon2023,Lau2025}.

\citet{De2023} discovered the infrared and optical transient \event{} and interpreted it as a star-planet merger, providing the first real-time observation of such an event\@. \citet{Lau2025a} observed the merger remnant roughly two years later. The star-planet merger interpretation of this transient relies in part on its qualitative similarity to luminous red novae (LRNe), a transient class associated with the mergers of two stars \citep{Tylenda2011,Ivanova2013}. However, the radiated energy and ejecta mass of \event{} are smaller by a factor of roughly a hundred, suggesting a merger between a star and a much smaller companion.

Here we model the evolution of \event{}. Section~\ref{sec:observations} summarizes the observations of \event{} \citep{De2023,Lau2025a}, including the progenitor system and the outburst. Section~\ref{sec:pre_merger} discusses the evolution of the system before the merger, which we model as tidal decay of the planetary orbit. Section~\ref{sec:merger} discusses the evolution of the system once the star and planet come into contact. The planet shocks and ejects stellar material near the surface; this interaction gradually intensifies until the planet is fully immersed in the star. Section~\ref{sec:transient} compares \event{} to stellar mergers and discusses the potential radiative processes responsible for the transient. Section~\ref{sec:energetics} combines the results of previous sections to place energetic constraints on the mass of the planet.%

\section{Summary of observations}\label{sec:observations}
Observations of \event{} include (i) archival near-infrared images of the progenitor star at the \qty{-12}{\year} epoch (defined with respect to the peak of the optical transient), (ii) mid-infrared photometry starting at the \qty{-244}{\day} epoch, with a cadence of a few months, (iii) optical photometry for \( \qty{-30}{\day} \lesssim t \lesssim \qty{150}{\day} \), with a cadence of a few days, and (iv) infrared spectroscopy of the merger remnant at the \qty{830}{\day} epoch.%

\event{} is located in the galactic disk, a distance
\begin{equation}
d \approx \qty{4}{\kilo\parsec}
\end{equation}
from Earth. Figure~1 from \citet{De2023} shows the optical and infrared light curves. Their Figure~2 shows the bolometric properties (see also Figure~\ref{fig:bolometric_properties}). We describe the observations in more detail below.

\subsection{Progenitor}\label{sec:observations:progenitor}
The progenitor star appears in \(H \) and \(K \) images from the United Kingdom Infrared Telescope (UKIRT) survey of the galactic plane \citep{Lawrence2007}, taken at the \( - \qty{12}{\year} \) epoch. Multiple surveys covered the location of the progenitor, including PanSTARRS1 \citep{Chambers2016}, Gaia \citep{GaiaCollaboration2016,GaiaCollaboration2021}, and POSS-II \citep{Reid1991}, but none of them detected it. The position of the progenitor in the HR diagram, as determined by the UKIRT images, is consistent with evolutionary tracks in the mass range \(0.8\lesssim\mstar/\unit{\solarmass}\lesssim1.5 \) \citep{De2023}, although there are significant photometric errors. The \( \qty{1}{\solarmass} \) track is consistent with the observed progenitor for stellar radii between \( \qty{1}{\solarradius} \) and \( \qty{4}{\solarradius} \), suggesting a star on or close to the main sequence\@. \citet{Lau2025a} observed the source at the \qty{830}{\day} epoch, and estimated an intrinsic stellar luminosity \( \lstar \approx \qty{0.3}{\lsun} \), corresponding to a \( \approx \qty{0.7}{\solarmass} \) main-sequence star. However, it is possible that dust is obscuring a slightly more luminous (i.e., more massive) star. The planet is not observable in the archival image; \citet{De2023} inferred its presence from the properties of the transient, so there are no direct constraints on its properties.

We will use a sunlike star (\qty{1}{\solarmass}, \qty{1}{\solarradius}) in our analysis. We present our calculations including the dependence on stellar mass and radius, so that they can be scaled to different stars. Some of our analysis requires knowledge of the internal structure of the star, which we calculate using the \verb|1M_pre_ms_to_wd| inlist from the modules for experiments in stellar astrophysics \citep[MESA;][and additional citations in the software section]{Paxton2011}.

\subsection{Pre-merger dust and infrared brightening}
The source began to form dust and gradually brighten in the infrared around \( \approx \qty{200}{\day} \) before the peak of the optical transient \citep[see the WISE data points in Figure~1 from][]{De2023}\@. \citet{De2023} fit the infrared spectral energy distribution (SED) to estimate the mass of the dust around the source. Assuming a dust-to-gas mass ratio \( 10^{-2} \), they estimated a total (dust plus gas) ejecta mass of
\begin{gather}
\mej\lp t=\qty{-244}{\day}\rp \approx \qty{2.8e-5}{\solarmass}\lp\frac{d}{\qty{4}{\kilo\parsec}}\rp^2,\\
\mej\lp t=\qty{-44}{\day}\rp \approx \qty{e-4}{\solarmass}\lp\frac{d}{\qty{4}{\kilo\parsec}}\rp^2
\end{gather}
at the \( t = \qty{-244}{\day} \) and \( t = \qty{-44}{\day} \) epochs, respectively, where \(t=0\) corresponds to the peak of the optical transient.

\subsection{Optical transient}\label{sec:observations:optical_transient}
The optical luminosity increased significantly over the course of approximately ten days, reaching a peak at \(\qty{1.3e35}{\erg\per\second} \) that lasted about \(\qty{25}{\day} \). After that, the transient faded by about an order of magnitude over \( \qty{150}{\day} \) (Figure~\ref{fig:bolometric_properties}). The total radiated energy over this period was
\begin{equation}
E_\mathrm{rad}\approx\qty{6e41}{\erg}\lp \frac{d}{\qty{4}{\kilo\parsec}}\rp^2.
\end{equation}
\citet{De2023} estimated the bolometric properties from the optical peak to \( \approx \qty{120}{\day} \) post-peak by fitting a blackbody function to the ZTF and ATLAS photometry; see Section 13 of their supplemental information for more details. They also estimated the properties of the dust that formed around the remnant as a result of the merger using SEDs at the \qty{120}{\day} and \qty{320}{\day} epochs (see their extended data table 3). They found that the dust was expanding at a speed \( \approx \qty{35}{\kilo\meter\per\second} \).

\citet{Lau2025a} reanalyzed the optical and infrared SED of \event{} SED at \qty{320}{\day} and arrived at new estimates for the properties of the dust surrounding the remnant. Their model of the dust assumes that it is distributed in a spherical shell. The parameters of this shell, such as the inner radius, temperature, and optical depth are free parameters (see their table 3 for their best-fit values). Assuming a dust-to-gas mass ratio \( 10^{-2} \), they estimated a total (dust plus gas) ejecta mass of
\begin{equation}
\log_{10}\lp \frac{\mej\lp t=\qty{320}{\day}\rp}{\unit{\solarmass}}\rp=-3.89_{-3.21}^{+0.29}.\label{eq:m_ej_obs_320}
\end{equation}

\subsection{Remnant at \qty{830}{\day}}
\citet{Lau2025a} observed \event{} at \qty{830}{\day} using JWST and Gemini North. From the SED, they estimated \( \approx \qty{e-9}{\solarmass} \) of warm (\(\approx\qty{700}{\kelvin}\)) gas around the remnant. They suggested that this gas forms an accretion disk around the remnant, based on their detection of \( ^{12}\mathrm{CO} \) and \( \mathrm{Br}\alpha \) emission. They also constrained the properties of the progenitor and reanalyzed the \qty{320}{\day} SED, as we discussed above.

\section{Pre-merger evolution}\label{sec:pre_merger}

We will discuss the evolution of the system before the star and the planet come into contact, and determine the observability of the star-planet interaction at the time of the archival images of the progenitor (\(\qty{12}{\year} \) before the transient).

The presence of dust in the months prior to the main outburst suggests that the merger was the culmination of an escalating interaction on a timescale of \( \gtrsim \) months. Therefore, we focus on tidal dissipation as the likely mechanism that led to the merger. Orbital decay has been observed for WASP\nobreakdash-12~b \citep{Maciejewski2016,Patra2017,Maciejewski2018,Bailey2019,Yee2020,Turner2021,Leonardi2024} and Kepler\nobreakdash-1658~b \citep{Vissapragada2022}, although the exact mechanism of tidal dissipation which would explain the observed rates of decay remains unclear \citep{Weinberg2024,Barker2024,Millholland2025}. We will begin by modeling the pre-merger orbital evolution of the system subject to tidal dissipation.

\subsection{Tidal orbital decay}
The orbital period of the close-in (orbital separation \( \approx \) stellar radius) planet is
\begin{equation}
\begin{split}
\porb=&2\pi\lp G\mstar/a^3\rp^{-1/2}\\
\approx&\qty{2.8}{\hour}
\lp\mstar/\unit{\solarmass}\rp^{-1/2}
\lp a/\unit{\solarradius}\rp^{3/2},
\end{split}
\end{equation}
where \(G \) is the gravitational constant, \(\mstar \) is the mass of the star, and \(a \) is the orbital separation. The orbital energy is, assuming a circular orbit,
\begin{equation}
\begin{split}
\eorb=&-\frac{G\mstar\mplanet}{2a}\\
=&-\qty{1.8e45}{\erg}
\lp\frac{\mstar}{\unit{\solarmass}}\rp
\lp\frac{\mplanet}{\unit{\jupitermass}}\rp
\lp\frac{a}{\unit{\solarradius}}\rp^{-1},
\end{split}
\end{equation}
where \(\mplanet \) is the mass of the planet and \( \unit{\jupitermass} \) is the mass of Jupiter. Tides dissipate orbital energy at a rate
\begin{equation}
\begin{split}
\edottide =& -
\frac{9}{2\qstar} \lp\frac{\mplanet}{\mstar}\rp
\lp\frac{\rstar}{a}\rp^5
n \eorb \\
=&
\qty{13}{\solarluminosity}
\lp\frac{\mplanet}{\unit{\jupitermass}}\rp^2
\lp\frac{\mstar}{\unit{\solarmass}}\rp^{1/2}
\lp\frac{\rstar}{\unit{\solarradius}}\rp^5
\lp\frac{\qstar}{10^5}\rp^{-1}\\&
\times\lp\frac{a}{\unit{\solarradius}}\rp^{-15/2},
\label{eq:ltide}
\end{split}
\end{equation}
where \(n\equiv2\pi / \period \) is the orbital mean motion and \qstar{} is the modified tidal quality factor of the star, which parametrizes the strength of the mechanism responsible for the tidal dissipation. A large \qstar{} corresponds to weak dissipation, and a small \qstar{} to strong dissipation \citep[for a review of tidal dissipation in stars and giant planets, see][]{Ogilvie2014}. The quality factor likely varies by several orders of magnitude depending on the properties of the star, planet, and the orbit \citep{Barker2020}. Here, we keep \qstar{} as a free parameter with a power-law dependence on orbital period,
\begin{equation}
\qstar\propto\porb^\alpha.
\end{equation}
The timescale of orbital decay is
\begin{equation}
\ttide=-\int_0^{a}\frac{da'}{\dot{a}'}=\frac{2}{13+3\alpha}\frac{\eorb}{\edottide}.\label{eq:t_decay}
\end{equation}

We also will consider also consider a range of constant quality factors between \( 10^4 \) and \( 10^8 \)\@. \citet{Penev2018} observationally constrained the tidal quality factor using the observed rotation rates of a sample of hot-Jupiter host stars\@. \citet{Millholland2025} did a similar experiment using the steady-state distribution of planetary orbital parameters. Under the assumption that the planets had tidally increased the rotation rate of their stars on a timescale equal to the age of the stars, \citet{Penev2018} found a tidal quality factor
\begin{equation}
\qpenev =\max\left[
3.52\times10^6\lp\frac{\porb}{\qty{1}{\day}}\rp^{-3.1}, 10^5
\right],
\label{eq:q_penev}
\end{equation}
which is broadly consistent with the later findings of \citet{Millholland2025} (see their Figure 7).
We also consider this empirical tidal quality factor in our analysis, noting that population-level patterns do not necessarily predict the behavior of individual systems because the tidal quality factor might vary by orders of magnitude across systems.%

\subsection{Observability of tidal heating}
Here we determine whether tidal heating changes the appearance of the star before the merger. There are two requirements for tidal heating to be observable: one, that the tidal energy deposition rate be comparable to the intrinsic stellar luminosity; and two, that the star respond to the tidal energy dissipated in its interior on a timescale shorter than the orbital decay time. For quality factors \(\qstar\lesssim10^6\lp\mplanet/\unit{\jupitermass}\rp^2 \), the tidal energy dissipation rate is greater than the stellar luminosity when \( a \approx \rstar \).

These requirements relate not only to the magnitude of the heating, but also to its location within the star \citep{Podsiadlowski1996}. To illustrate the latter, consider a process which dissipates energy close to the center of the star (e.g., internal gravity waves). In order to expand, the heated material must share its energy with the layers above. As a result, heating will change the (quasi)hydrostatic stellar structure on the timescale on which the heating changes the total energy of the star,
\begin{equation}
\tau_\mathrm{heat,global}=\edottide^{-1}\int_{0}^{\mstar} c_p T\,dM \approx \frac{G \mstar^2}{2\rstar\edottide},
\end{equation}
where \( c_p \) is the specific heat capacity, \( T \) is the temperature, and we approximated the internal energy of the star as \( 0.5 G \mstar^2 / \rstar \). This timescale is always longer than the timescale of orbital decay,
\begin{equation}
\approx \frac{G \mstar\mplanet}{a\edottide},
\end{equation}
because the energy in the orbit is smaller than the energy in the star by a factor \(\approx\lp\mplanet/\mstar\rp\ll1 \). Therefore, central heating is unlikely to be observable before the merger.

In contrast, consider a process that heats the outermost \(M_\mathrm{heat} \) of the star (e.g., turbulent dissipation of the equilibrium tide). The timescale over which this region will expand is the local heating time
\begin{equation}
\theat\approx f_\mathrm{}\frac{G \mstar M_\mathrm{heat}}{2\rstar\edottide},\label{eq:t_heat}
\end{equation}
where \( f \lesssim 1 \) is a numerical factor accounting for the fact that \( c_p T < G \mstar / \rstar \) in the outer regions of the star. If the energy is deposited close to the surface (i.e., \(M_\mathrm{heat}\ll\mstar \)), the local heating time can be much shorter than the global one.
We can combine equations~\eqref{eq:t_decay}~and~\eqref{eq:t_heat} to relate the thermal and orbital decay timescales,
\begin{equation}\label{eq:tdecay_tthermal}
\ttide\approx f^{-1}\frac{2}{13+3\alpha}\frac{\mplanet}{M_\mathrm{heat}}\frac{\rstar}{a}\theat.
\end{equation}
Equation~\eqref{eq:ltide} shows that \( \edottide/\lstar \propto \mplanet^2 \), and equation~\eqref{eq:tdecay_tthermal} shows that \( \ttide/\theat \propto \mplanet \), so heating from more massive planets is more likely to be observable.

\begin{figure}[t!]
\centering
\includegraphics[width=\columnwidth]{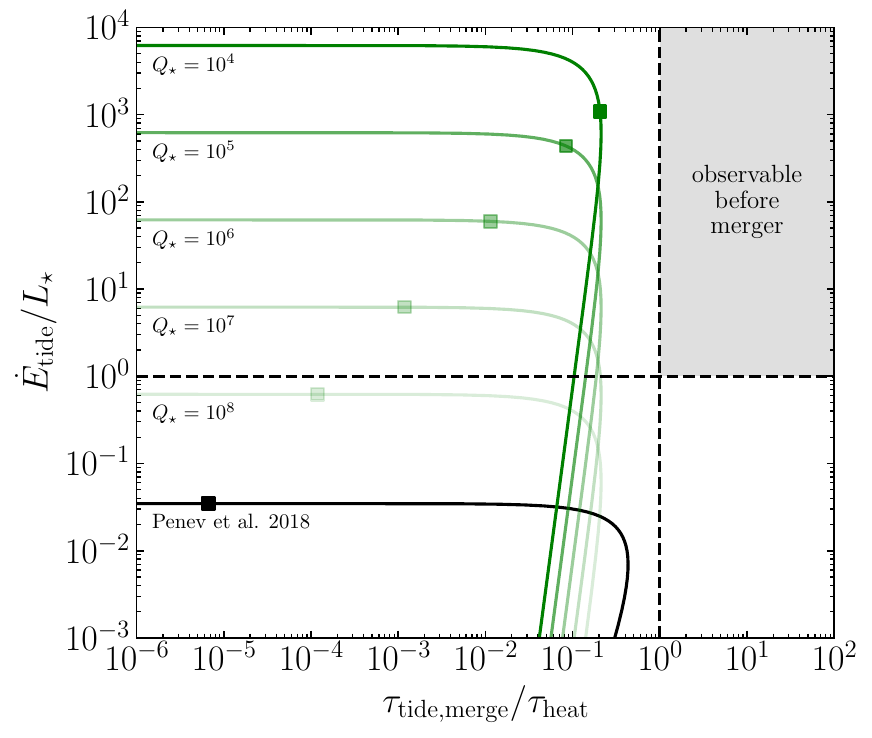}
\caption{Tidal evolution tracks for a sunlike star with a \( \qty{10}{\jupitermass} \) companion. The abscissa shows the ratio of the orbital decay time to the thermal time at the location of energy deposition (see equations~\ref{eq:tdecay_tthermal}~and~\ref{eq:tau_tide_merge}); the ratio must be greater than unity for the deposited energy to be observable before the merger. The ordinate shows the ratio of the tidal luminosity to the intrinsic stellar luminosity (see equation~\ref{eq:ltide}); the ratio must be greater than unity for the tidal energy deposition to be significant. Each line corresponds to a different tidal quality factor. The squares mark the \(\qty{-12}{\year} \) epoch at which the archival images of the progenitor were taken. When computing the thermal time, we assume energy is deposited in the outermost \( \qty{2e-2}{\solarmass} \) of the star. None of these tidal quality factors yield observable tidal heating before the merger (upper right region of the figure), suggesting that the archival image was unaffected by the star-planet interaction.}\label{fig:tides_observability}
\end{figure}

Figure~\ref{fig:tides_observability} shows tidal evolution tracks in the plane defined by \(\edottide/\lstar \) and \(\tau_\mathrm{tide,merge}/\theat \), where
\begin{equation}
\tau_\mathrm{tide,merge}\lp a\rp\equiv \ttide\left(a\right)-\ttide\left(\rstar+\rplanet\right)\label{eq:tau_tide_merge}
\end{equation}
is the tidal orbital decay time to the orbital separation at which the star and the planet come into contact. We use a sunlike star with a \( \qty{10}{\jupitermass} \) companion. We assume \( f = 0.2 \) and \( M_\mathrm{heat}=\qty{2e-2}{\solarmass} \), corresponding to a mechanism that deposits heat near the base of the outer convective zone. Initially, \( \tau_\mathrm{tide,merge}/\theat\propto a^{-1} \), as in equation~\eqref{eq:tdecay_tthermal}. However, as the orbital separation approaches \( \rstar \), \( \tau_\mathrm{tide,merge} \) approaches zero while \( \theat \) approaches a constant positive value, so all curves bend toward the left of the plot as \(a\) approaches \( \rstar \).

Figure~\ref{fig:tides_observability} shows that tidal heating of a sunlike star by even a massive \( \qty{10}{\jupitermass} \) planet does not significantly affect the stellar structure before the merger, and therefore was likely unobservable at the time of the archival image (\( \qty{12}{\year} \) before the merger). The tidal heating of the star could be observable if the heated region is closer to the surface (\( M_\mathrm{heat} \ll \qty{2e-2} \solarmass \)) because \( \ttide/\theat\propto M_\mathrm{heat}^{-1} \). In Section~\ref{sec:transient} we will argue that a qualitatively similar process---the transfer of orbital energy from the orbit into the outermost layers of the star---is responsible for the transient.

\subsection{Planetary structure}
As the orbital separation decreases, the planet experiences increased irradiation and tidal forces.
Irradiation might inflate the planet \citep{Hartman2016,Komacek2017,Thorngren2021}; if 1\% of the stellar flux reaches the center of the planet as heat, a warm Jupiter could expand in radius by up to a factor of a few as its star evolves off the main sequence \citep{Lopez2016}. This expansion would make the planet more vulnerable to photoevaporation \citep[e.g.,][]{MurrayClay2009} and tidal disruption.

Even if inflation is unimportant, the planetary mass-radius relation implies that some planets will overflow their Roche lobe above the surface of the star.
If the mass transfer is stable, the orbit of the planet expands to a period roughly determined by the mass of the core of the planet \citep{Valsecchi2014,Valsecchi2015,Jackson2016}, and subsequent tidal dissipation could again decrease the orbital separation. The combination of these effects could play a significant role in shaping exoplanet populations \citep[e.g.,][]{Lazovik2023,Thorngren2023}. On the other hand, if mass transfer is unstable, the star tidally disrupts and accretes the planet \citep[e.g.,][]{Faber2005,Guillochon2011,Liu2013}. Finally, if the planet is sufficiently dense, the planet can reach the stellar surface and fully merge with the star before being destroyed. This ``merger'' scenario is the focus of the following sections, although we briefly discuss tidal disruption in Section~\ref{sec:alt:tde}.%

\section{Merger}\label{sec:merger}
\subsection{Surface interaction}\label{sec:merger:surface_interaction}
When the surface of the star and the planet come into contact, the former exerts a drag force on the latter, further dissipating orbital energy. Initially, only the surface region of the star is in contact with the planet, so we refer to this phase as the surface interaction. The strength of the drag force depends on the relative speed between the planet and the stellar surface. Before the merger, tidal dissipation transfers angular momentum from the orbit of the planet to the star, increasing its rotation rate. If the orbital and stellar rotation frequencies are equal, the system is said to be in corotation. We can estimate the extent of corotation at the onset of the surface interaction by equating the change in orbital angular momentum to the change in the spin angular momentum of the star. The change in orbital angular momentum from an initial separation \( a_0 \) to \rstar{} is transferred to the star, i.e., \( I_\star \omega_\mathrm{\star,m} = \mplanet \lp a_0^2 \omega_0-\rstar^2\omega_\mathrm{Kep,m}\rp \), \( I_\star = \eta \mstar \rstar^2 \) is the moment of inertia of the star, \( \omega_{\star,\mathrm{m}} \) and \( \omega_\mathrm{Kep,m} \) are the stellar and Keplerian rotation frequencies at the onset of the merger, respectively, and \( a_0 \) and \( \omega_0 = \sqrt{G \mstar / a_0^3}\) are the initial separation of the planet and the orbital frequency at that separation, respectively. We find
\begin{equation}
\begin{split}
\frac{\omega_{\star,\mathrm{m}}}{\omega_\mathrm{Kep,m}}&=\eta^{-1}\frac{\mplanet}{\mstar}\lp\sqrt{\frac{a_0}{\rstar}}-1\rp \\
&\approx10^{-2}\lp\frac{\eta}{0.08}\rp^{-1}\lp\frac{\mplanet}{10^{-3}\mstar}\rp\lp\sqrt{\frac{a_0}{\rstar}}-1\rp.
\end{split}
\end{equation}
The rotation periods of hot-Jupiter host stars are typically of order weeks \citep[see Table 1 from][]{TejadaArevalo2021}, much longer than the Keplerian orbital period at their surface. In that sense, these host stars rotate slowly, and their initial rotation rates are negligible. For a typical hot Jupiter with \( a_0 = \qty{0.05}{\astronomicalunit} \approx 10\rstar \) this equation shows that the rotation rate at the onset of the surface interaction will be much smaller than the Keplerian one \citep[see also][]{Soker2006,Privitera2016}. We therefore approximate the relative speed between the planet and the stellar material as the Keplerian speed,

\begin{equation}
\vorb \approx \sqrt{G \mstar / a}
\end{equation}

\begin{figure*}[t!]
\centering
\includegraphics[width=\columnwidth]{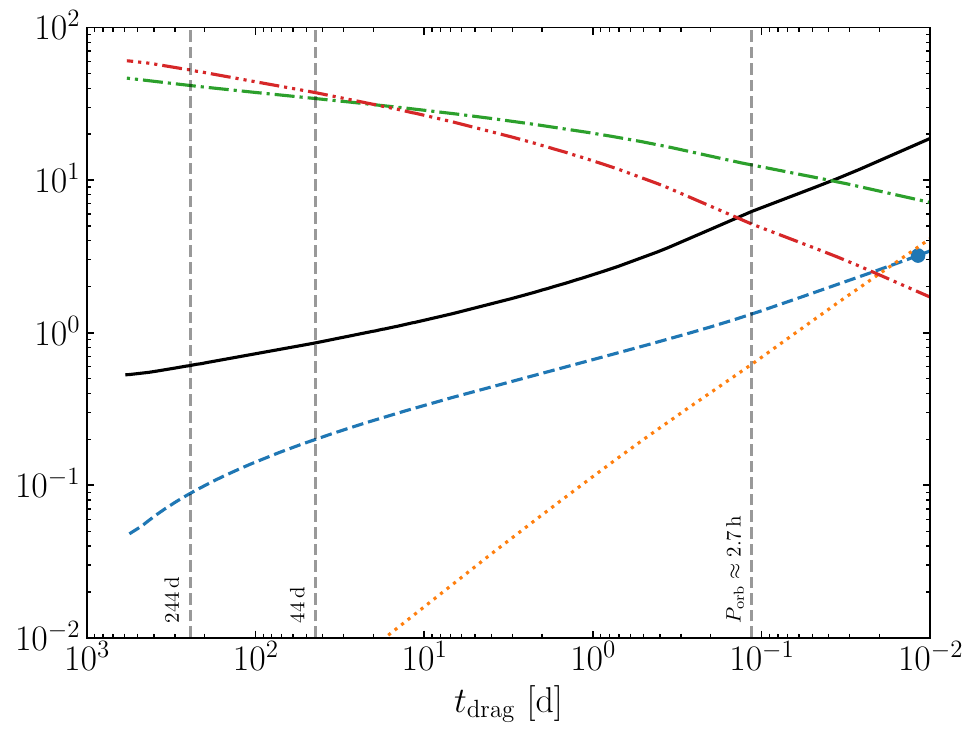}
\includegraphics[width=\columnwidth]{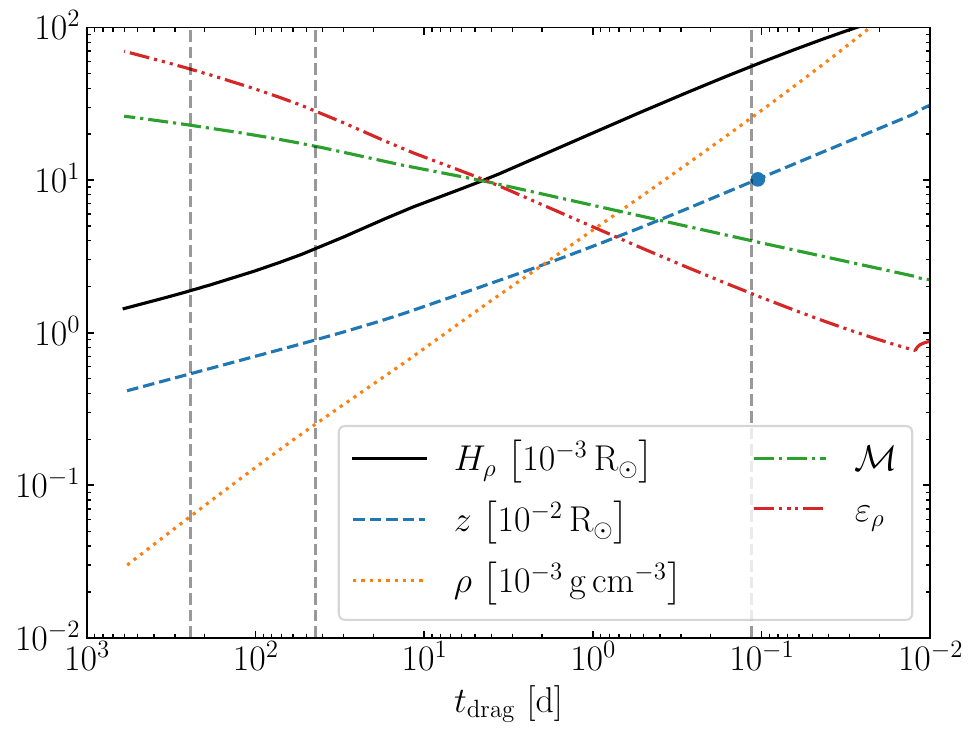}
\caption{Important quantities during the merger between a sunlike star and a Neptune (\( \qty{15}{\earthmass}, \qty{3.5}{\earthradius} \), left panel) or a giant planet (\( \qty{10}{\jupitermass} \), \( \qty{1}{\jupiterradius} \), right panel), as a function of the timescale of orbital decay as a result of drag. Vertical dashed lines show the two epochs at which pre-merger constraints exist, as well as the orbital period at the surface of the star. The plots show the Mach number of the planet \(\mathcal{M} \equiv \vorb / c_\mathrm{s}\), \(\epsrho\equiv \) number of density scale heights across the planet, the depth \( z \equiv \rstar - r \) (with a circle indicating the location where the depth equals twice the radius of the planet), the stellar density \( \rho \), and the density scale height \hrho{}. The motion of the planet is always supersonic, leading to shocks. During the surface interaction, the flow is strongly stratified at the scale of the planet (\( \epsrho > 1 \)).}\label{fig:dimensionless_quantities}
\end{figure*}

The planet experiences drag as a result of both ram pressure and gravitational interactions with the stellar material. The ratio between these two drag forces is roughly the ratio between the geometrical (\(\pi\rplanet^2 \)) and gravitational (\(\pi R_a^2 \), where \(R_a\equiv2G\mplanet/\vorb^2 \)) \crosssection{}s of the planet,
\begin{equation}
\begin{split}\label{eq:dragratio}
\lp\frac{\rplanet}{R_a}\rp^2
\approx&3\times10^3
\lp\frac{\mstar}{\unit{\solarmass}}\rp^{2}
\lp\frac{\mplanet}{\unit{\jupitermass}}\rp^{-2}
\lp\frac{\rplanet}{\unit{\jupiterradius}}\rp^2\\
&\times\lp\frac{a}{\unit{\solarradius}}\rp^{-2},
\end{split}
\end{equation}
For the star-planet combinations relevant to \event{}, ram pressure drag dominates \citep[see also Figure 2 from][]{Yarza2023}. The ratio between ram pressure drag and gravitational drag increases as the planet orbit decays because the orbital speed increases, shrinking the gravitational \crosssection{}. We therefore consider only ram pressure drag hereafter.

The magnitude of the drag force depends on the stratification of the stellar density profile; if the density scale height is much smaller than the size of the planet, its effective \crosssection{} is the area within roughly a scale height of its substellar point \citep{Metzger2012}. On the other hand, when the density scale height is much larger than the radius of the planet, the density is approximately constant across the surface of the planet, and the \crosssection{} of the planet is its standard geometrical \crosssection{} \(\pi\rplanet^2 \). We can quantify stratification using the number of density scale heights across a planet radius, \(\epsrho\equiv\rplanet/\hrho \). When \(\epsrho=0 \), the density is constant, and when \(\epsrho\gg1 \), the density is strongly stratified on the scale of the planet. We will now estimate the properties of the star near its surface, including \epsrho{}. We approximate the temperature near the surface as that of a thin adiabatic gas in hydrostatic equilibrium \citep{Yamazaki2017},
\begin{equation}
\begin{split}
T&\approx\frac{\gamma-1}{\gamma}\frac{g z \mu}{k}\\
&\approx\qty{5.5e4}{\kelvin}\lp\frac{\mstar}{\unit{\solarmass}}\rp\lp\frac{\rstar}{\unit{\solarradius}}\rp^{-2}\lp\frac{z}{\qty{e-2}{\solarradius}}\rp,
\end{split}
\end{equation}
where
\begin{equation}
z \equiv \rstar - a
\end{equation}
is the depth, \(g=G \mstar/\rstar^2 \) is the gravitational acceleration, \( k \) is the Boltzmann constant, \( \mu \) is the mean molecular weight (we used that of a fully ionized gas with hydrogen and metal mass fractions 0.74 and 0.02, respectively) and \( \gamma \approx 5 / 3 \) is the adiabatic index. The scale height is \(\hrho\approx \cs^2/g \), where
\begin{equation}
\begin{split}
\cs=&\sqrt{\frac{\gamma k T}{\mu}}\approx\sqrt{\frac{\gamma-1}{\gamma}\frac{z}{\rstar}\frac{G\mstar}{\rstar}}\\
\approx&\qty{2.8e6}{\centi\meter\per\second}\lp\frac{\mstar}{\unit{\solarmass}}\rp^{1/2}\lp\frac{\rstar}{\unit{\solarradius}}\rp^{-1}\\
&\times\lp\frac{z}{\qty{e-2}{\solarradius}}\rp^{1/2}
\label{eq:cs_approx}
\end{split}
\end{equation}
is the speed of sound. We find
\begin{gather}
\hrho\approx\frac{\gamma-1}{\gamma}z\approx\qty{4e-3}{\solarradius}\lp\frac{z}{\qty{e-2}{\unit{\solarradius}}}\rp,\\
\epsrho\equiv\frac{\rplanet}{\hrho}\approx\frac{\gamma}{\gamma-1}\frac{\rplanet}{z}\approx25\lp\frac{\rplanet}{\unit{\jupiterradius}}\rp\lp\frac{z}{\qty{e-2}{\unit{\solarradius}}}\rp^{-1}.
\end{gather}

These estimates show that the stellar density is strongly stratified during the surface interaction. In Appendix~\ref{sec:appendix:drag} we derive the \crosssection{} of the planet as a function of \epsrho{} \citep[see also][]{Metzger2012}. The drag force is
\begin{equation}
F_\mathrm{d}\approx\rho \vorb^2\sigma\simeq\sqrt{2\pi} \rho \vorb^2 \hrho^{3/2} \rplanet^{1/2},\label{eq:drag_high_epsrho}
\end{equation}
where \( \rho \) is the density at the substellar point of the planet, and \(\sigma\simeq\sqrt{2\pi}\hrho^{3/2}\rplanet^{1/2} \) is the \crosssection{} of the planet in the \(\epsrho\gg1 \) limit. Given this drag force, the characteristic timescale of orbital decay is
\begin{equation}
\begin{split}
\tdrag\approx\frac{\hrho}{a}\frac{\eorb}{\dot{E}_\mathrm{drag}}=&\qty{16}{\hour}
\lp\frac{\mplanet}{\unit{\jupitermass}}\rp
\lp\frac{\rho}{\qty{e-3}{\gram\per\cubic\centi\meter}}\rp^{-1}\\
&\times\lp\frac{\rstar}{\unit{\solarradius}}\rp^{-1/2}
\lp\frac{\hrho}{\qty{e-2}{\solarradius}}\rp^{-1/2}\\
&\times\lp\frac{\rplanet}{\unit{\jupiterradius}}\rp^{-1/2}
\lp\frac{\mstar}{\unit{\solarmass}}\rp^{-1/2},\label{eq:tdrag}
\end{split}
\end{equation}
where \( \dot{E}_\mathrm{drag} = F_\mathrm{d}\vorb \) is the rate of energy dissipation as a result of drag, and the factor of \( \hrho / a \) accounts for the fact that the orbit need only decay by approximately a scale height for the drag to increase significantly.
When the star and the planet just come into contact (\( a < \rstar + \rplanet \)), the density is so low (\( \rho \ll  \qty{e-3}{\gram\per\cubic\centi\meter} \)) that tides will still dominate. The timescale of orbital decay from tides near the surface is
\begin{equation}
\begin{split}
\left.\lp\frac{\hrho}{a}\rp\ttide\right|_{a=\rstar}=&\frac{4\qstar}{117}\frac{\hrho}{\rstar}\frac{\mstar}{\mplanet}\sqrt{\frac{\rstar^3}{G \mstar}}\\
\approx&\qty{18}{\year}\lp\frac{\hrho}{10^{-2}\rstar}\rp\lp\frac{\qstar}{10^6}\rp \\
&\times\lp\frac{\mstar}{\unit{\solarmass}}\rp^{1/2}\lp\frac{\mplanet}{\unit{\jupitermass}}\rp^{-1} \\
&\times\lp\frac{\rstar}{\unit{\solarradius}}\rp^{3/2}\label{eq:local_ttide_at_surface},
\end{split}
\end{equation}
where we used equation~\eqref{eq:t_decay} for the tidal decay timescale. The factor of \( \hrho / a \) appears for the same reason as in equation~\eqref{eq:tdrag}. Equation~\eqref{eq:local_ttide_at_surface} implies that for low quality factors or high planet masses, the orbital decay timescale at the surface could be months, and that the star and planet might not be in contact at the \qty{-244}{\day} epoch. If we instead assume the \citet{Penev2018} tidal quality factor, then, regardless of planet mass, tides are so inefficient at short orbital periods that drag dominates during the observed pre-merger epochs. Given the uncertainty in the tidal quality factor, and the difficulty in quantifying the drag-tides interaction without detailed models of the evolution of the stellar atmosphere, we assume drag dominate the orbital evolution once the star and planet come into contact, and use equation~\eqref{eq:tdrag} as an estimate of the timescale of orbital decay.

Figure~\ref{fig:dimensionless_quantities} shows important quantities as a function of the drag decay time, equation~\eqref{eq:tdrag}. We computed these quantities using the MESA stellar model (which we described in Section~\ref{sec:observations:progenitor}); the analytical estimates from the previous section give similar values. The left and right panels correspond to a Neptune and a gas giant, respectively. For most of the surface interaction, the orbital separation changes by a fractionally small amount (i.e., \( a \approx \rstar \)), so the orbital period is approximately equal to the orbital period at the stellar surface, which we show as a vertical dashed line. A blue line shows the depth of the planet within the envelope, with a blue dot showing when the planet becomes fully immersed. The orbital decay becomes dynamical (i.e., \( \tdrag = \porb \)) during the surface interaction (i.e., when \( z \lesssim 2\rplanet \)), at which point the planet plunges into the stellar interior. The stellar density is strongly stratified (red line, \( \epsrho \gtrsim 1 \)). We will discuss the other quantities shown in the figure when we discuss the physical processes to which they are relevant.%

\subsection{Ejecta}\label{sec:merger:ejecta}
We can use equation~\eqref{eq:cs_approx} to estimate the Mach number of the motion of the planet in the star,
\begin{equation}
\begin{split}
\mathcal{M}\equiv&\frac{\vorb
}{\cs}\approx\sqrt{\frac{\gamma}{\gamma-1}\frac{\rstar}{z}}\\
&\approx16\lp\frac{z}{\qty{e-2}{\solarradius}}\rp^{-1/2}\lp\frac{\rstar}{\unit{\solarradius}}\rp^{1/2}.
\end{split}\label{eq:mach}
\end{equation}
\begin{figure}[t!]
\centering
\includegraphics[width=\columnwidth]{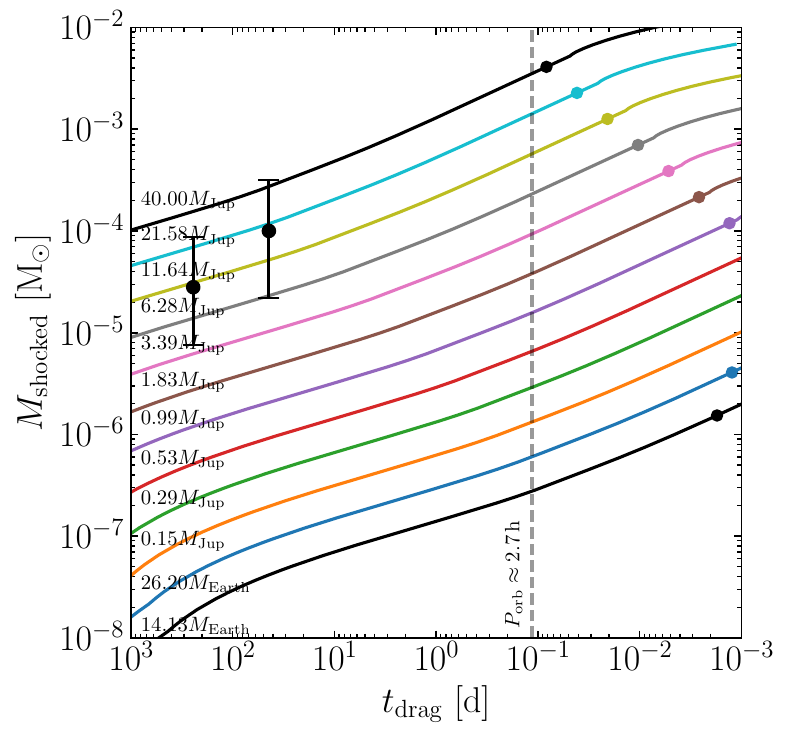}
\caption{Mass shocked by the planet as a function of the drag decay time, for different planet masses. Dots show the point at which the planet is fully immersed. A dashed vertical line shows the orbital period of the planet when the orbital separation equals the stellar radius. Black error bars show constraints for the ejecta mass at two pre-merger epochs. These constraints suggest a planet with \( \mplanet \gtrsim \qty{5}{\jupitermass} \) satisfy these constraints.}\label{fig:shocked_mass}
\end{figure}
Figure~\ref{fig:dimensionless_quantities} shows the Mach number computed using the stellar model, which agrees with the prediction from equation~\eqref{eq:mach} that the motion of the planet is always supersonic (\( \mathcal{M}>1 \)). Therefore, the planet will shock the stellar material. The observational consequences of these shocks depend on the location of the shocked material. If the material is close to the surface (e.g., during the surface interaction), the shocked material can form an outflow and escape without significant confinement from stellar material above it. In contrast, if the shocked material is deep in the interior (e.g., once the planet is fully immersed), the energy of the shocks will be shared with the layers closer to the surface, and the planet can be more reasonably approximated as a heat source in the stellar interior. %
While the details of ejecta formation depend on the hydrodynamics of the interaction, we use full immersion as the approximate condition at which the efficiency of ejecta formation decreases significantly.

The energy deposited close to the surface can escape more easily, so it is observable on shorter timescales. This near-surface energy deposition can more easily produce ejecta either dynamically or as a wind, depending on the duration, amplitude, and depth of the energy deposition. The energy deposited deeper in the star is likely to escape only on the longer \kh{} time of the envelope, and might not be observable on the timescale of the \event{} transient. For this reason, we argue that the surface interaction is responsible for the transient \citep[see also][]{Soker2006}.%

The change in orbital energy as the orbit of the planet decays from \( a = \rstar \) to \( a = \rstar - z \) is
\begin{equation}
\begin{split}
\Delta \eorb \approx& \left.\frac{d\eorb}{da}\right|_{a=\rstar}z = \eorb z/\rstar,\\
=&\qty{1.86e44}{\erg}
\lp\frac{\mplanet}{\unit{\jupitermass}}\rp
\lp\frac{\mstar}{\unit{\solarmass}}\rp
\lp\frac{z}{\unit{\jupiterradius}}\rp \\
&\times
\lp\frac{\rstar}{\unit{\solarradius}}\rp^{-2},
\end{split}
\end{equation}
where we let \( a\approx\rstar \). The planet will impart an average specific energy of order \( \vorb ^2\) on the shocked material. We can estimate the amount of mass that is shocked during the surface interaction as
\begin{equation}
\begin{split}
M_\mathrm{shocked}\approx&\Delta\eorb/\vorb^2=\frac{1}{2}\mplanet\frac{z}{\rstar} \\
\approx &\qty{4.9e-5}{\solarmass}
\lp\frac{\mplanet}{\unit{\jupitermass}}\rp
\lp\frac{z}{\unit{\jupiterradius}}\rp
\lp\frac{\rstar}{\unit{\solarradius}}\rp^{-1}.
\end{split}\label{eq:shocked_mass}
\end{equation}
A fraction of this shocked mass will become unbound. The shocked mass is much smaller than the mass of the planet because during the surface interaction the planet transfers into the envelope only a fraction
\( \rplanet / \rstar \approx 0.1\lp\rplanet/\unit{\jupiterradius}\rp\lp\rstar/\unit{\solarradius}\rp^{-1}\ll 1 \) of its orbital energy. If we assume that the efficiency of mass ejection decreases significantly when the planet becomes fully immersed, then we can estimate the total shocked mass by setting \( z = 2\rplanet \) and invert equation~\eqref{eq:shocked_mass} to find the minimum planet mass required to produce a given amount of ejecta,
\begin{equation}
\begin{split}
\mplanet>&\mej \frac{\rstar}{\rplanet}\\
=&\qty{1}{\jupitermass}\lp\frac{\mej}{\qty{e-4}{\solarmass}}\rp\lp\frac{\rstar}{\rsun}\rp\lp\frac{\rplanet}{\rjup}\rp^{-1}\label{eq:minimum_planet_mass_ejecta}
\end{split}
\end{equation}

Figure~\ref{fig:shocked_mass} shows the shocked mass as a function of the drag decay time for planets of different masses. The black error bars show estimates of the pre-merger ejecta mass \citep{De2023}. The vertical dashed line shows the orbital period at the surface of the star; the orbital decay becomes dynamical at that point. Circles show the point at which the planet is fully immersed in the star (i.e., \( z = 2\rplanet \)). The plot shows that a massive planet (\(\mplanet\gtrsim\qty{5}{\jupitermass}\)) is most consistent with the observations. We will examine the constraints on the planet mass in more detail in Section~\ref{sec:energetics}.

We have made several approximations in the arguments above and Figure~\ref{fig:shocked_mass}. The first one is the omission of tides, which we discussed in Section~\ref{sec:merger:surface_interaction}; if tides are important, then, at a given depth, drag is responsible for only a fraction of the dissipated orbital energy, and the corresponding shocked mass is smaller than if tides were negligible. The second is that we have not modeled the hydrodynamical interaction between the planet and the star. Finally, when plotting the observational constraints in Figure~\ref{fig:shocked_mass}, we have effectively assumed that the ejecta travels instantly to the dust condensation radius \( R_\mathrm{dust} \); in reality, the ejecta forming the dust must have been produced a time \( \approx \sqrt{R_\mathrm{dust}^3 / G \mstar} \) before the dust observation. The goal of this section has been to estimate the amount of energy available during the surface interaction; these estimates show that, if shocks during the surface interaction are the dominant mechanism producing ejecta, then only massive planets are consistent with the pre-merger observations.

\subsection{Destruction of the planet}
The planet experiences tidal forces and ram pressure in the star. These processes eventually destroy the planet. The evolution of the internal structure of the planet during the merger is uncertain, but is often estimated from order-of-magnitude arguments. The star tidally disrupts the planet approximately when the average densities of the planet and the mass enclosed by its orbit are equal. The ram pressure of the stellar gas disrupts the planet approximately when it equals the average binding energy per unit volume of the planet \citep{Jia2018}, i.e.,
\begin{equation}
\rho \vorb^2 = \bar{\rho}_\mathrm{p} v_\mathrm{esc,p}^2,
\end{equation}
where \( \bar{\rho}_\mathrm{p} \) is the average density of the planet and \( v_\mathrm{esc,p} = \lp 2G\mplanet/\rplanet \rp^{1/2} \) is the escape speed from its surface.

Hydrodynamical simulations suggest that the planet gradually loses mass to hydrodynamical ablation in the stellar envelope \citep[e.g.,][]{Murray1993,Sandquist1998,Passy2012,Lau2025}. The energetics of the debris is unclear; it can transfer some of its kinetic energy into the envelope, but it also gains thermal energy from the envelope because it is much colder \citep[see section 3.4 in][]{OConnor2023}. Given the uncertainties associated with these processes, we only note that the conditions for the destruction of both planets we consider in Figure~\ref{fig:dimensionless_quantities} are met only once they are fully immersed.%

\begin{figure}[t!]
\centering
\includegraphics[width=\columnwidth]{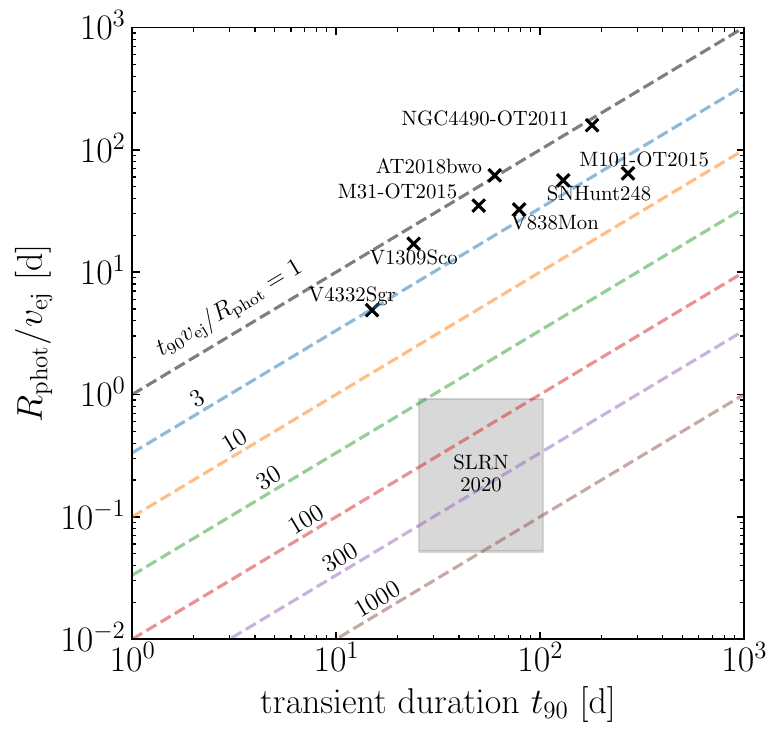}
\caption{Comparison between the time needed for ejecta at speed \(v_\mathrm{ej} \) to reach the observed photosphere (\(R_\mathrm{phot}/v_\mathrm{ej} \)) and the duration of the transient \(t_{90} \) (defined as the time since peak at which 90\% of the total radiated energy has been radiated). Scatter points show the observed properties of several stellar mergers. For most of them, these two timescales are within a factor of a few from each other\@. \event{} lies somewhere in the shaded region, depending on the definition of the duration of the transient (only the plateau \(\approx \qty{25}{\day} \) or the full light curve \( \approx\qty{100}{\day} \)) and on the assumed speed of the ejecta (ranging from the speed of the expanding inner dust shell \(\approx \qty{35}{\kilo\meter\per\second}\) to the escape velocity from a sunlike star \( \approx \qty{618}{\kilo\meter\per\second}\)). The duration of \event{} is much longer than the time it would take ejecta to reach the observed photosphere, so it is likely not powered by hydrogen recombination from a single episode of mass ejection.}\label{fig:transient_timescales}
\end{figure}

\section{The \event{} transient}\label{sec:transient}
\subsection{Comparison to luminous red novae}\label{sec:transient:lrn_comparison}
It is helpful to compare \event{} to LRNe, a transient class associated with stellar (i.e., star-star) mergers \citep{Tylenda2011,Ivanova2013}. Stellar and star-planet mergers have qualitatively similar light curves \citep[see the top panel of Figure 2c of][for a comparison between the light curves of \event{} and a few LRNe]{De2023}. These similarities motivate studying whether the same physical processes are responsible for producing their light curves and, more broadly, the extent to which star-planet mergers are ``scaled-down'' stellar mergers.

We begin by briefly summarizing the physical processes responsible for luminous red novae (LRNe) light curves. When the companion plunges into the stellar interior, it produces ejecta on a timescale comparable to the orbital period \citep{MacLeod2018a}. This ejecta has a distribution of speeds centered around \(\approx\vorb \) \citep[see, e.g., Figure~5 from][]{HutchinsonSmith2024}. This ejecta produces a light curve with two phases \citep{MacLeod2017a,Matsumoto2022}: (i) an initial peak, corresponding to the thermal emission of the high-velocity, low-mass ``tail'' of the ejecta energy distribution, and (ii) a longer plateau caused by the recombination of hydrogen in the bulk of the ejecta. The ejecta expands and cools until reaching the recombination temperature of hydrogen, \(\approx\qty{e4}{\kelvin} \). The radius at which the hydrogen recombines is approximately the radius of the photosphere because the opacity of atomic hydrogen is much smaller than that of ionized hydrogen.

The recombination radius is at least a factor of a few larger than the radius of the star, so ejecta must move at a significant fraction of the escape speed to reach the recombination radius. The slowest component of the ejecta to significantly contribute to the recombination transient therefore has speed \(\approx\vesc \). Therefore, the duration of the transient is roughly the time it takes for this marginally unbound component of the ejecta to reach the recombination radius, i.e., \(\approx R_\mathrm{phot}/\vesc \). The actual speed of the ejecta \( v_\mathrm{ej} \) might be different from \vesc{} depending on, e.g., the radius at which the ejecta is launched and the post-ejection evolution. Figure 2 from \citet{Matsumoto2022} shows that, for stellar mergers involving stars of roughly similar mass to \event{}, the observed ejecta speed is within a factor of a few from the surface escape speed of the star.

\begin{figure}[t!]
\centering
\includegraphics[width=\columnwidth]{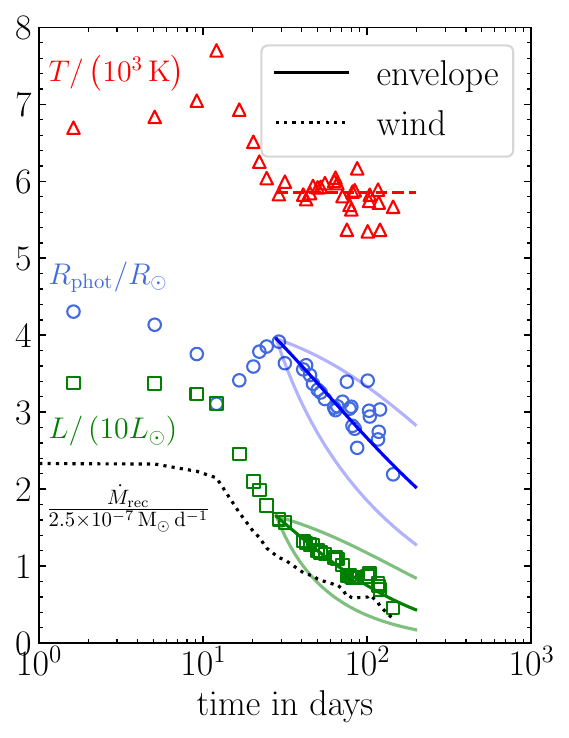}
\caption{Bolometric properties of \event{} as a function of time. Hollow points show the observations \citep{De2023}. Solid opaque lines show a model of a contracting envelope of mass \( \qty{1.1e-6}{\solarmass}\) around the merger remnant; semi-transparent lines show models with masses three times as small or as large. The horizontal dashed line is the constant effective temperature (\( \qty{5850}{\kelvin} \)) assumed in the envelope model. A dotted line shows the approximate required mass loss rate if the light curve were powered by the recombination of hydrogen in an outflow.}\label{fig:bolometric_properties}
\end{figure}

Figure~\ref{fig:transient_timescales} shows \(t_{90} \) (defined as the time since peak at which 90\% of the total radiated energy has been radiated) and \(R_\mathrm{phot}/v_\mathrm{ej} \) for a subset of known LRNe and for \event{}. These two timescales are within a factor of a few for all LRNe, in agreement with our rough estimate of the transient duration outlined above. We also show a shaded region for \event{}, accounting for uncertainties in the ejecta velocity and on the definition of transient duration appropriate in this context. As a lower limit for the speed, we use the expansion velocity of the inner radius of the dust shell \( v_\mathrm{ej}\approx\qty{35}{\kilo\meter\per\second} \), as determined from the SED at \qty{120}{\day} and \qty{320}{\day} \citep[Section~\ref{sec:observations:optical_transient} and][]{De2023}. As an upper limit, we use the escape speed from the surface of the star,
\begin{equation}
\vesc=\qty{618}{\kilo\meter\per\second}\lp \mstar/\unit{\solarmass}\rp^{1/2}\lp \rstar/\unit{\solarradius}\rp^{-1/2}.
\end{equation}
The two limits for the duration of the transient are the duration of the plateau (\( \approx\qty{25}{\day} \)) and of the entire light curve (\( \approx\qty{100}{\day} \)).

The figure shows that the duration of \event{} is much longer than \(R_\mathrm{phot}/v_\mathrm{ej} \). If the light curve were produced by the recombination of dynamically produced ejecta, the duration of the transient would be between
\begin{equation}
\frac{R_\mathrm{phot}}{\vesc}=\qty{1}{\hour}\lp\frac{R_\mathrm{phot}}{\qty{3.5}{\solarradius}}\rp
\lp\frac{\mstar}{\unit{\solarmass}}\rp^{-1/2}
\lp\frac{\rstar}{\unit{\solarradius}}\rp^{1/2}
\end{equation}
and
\begin{equation}
\frac{R_\mathrm{phot}}{\qty{35}{\kilo\meter\per\second}}\approx\qty{0.8}{\day}\lp\frac{R_\mathrm{phot}}{\qty{3.5}{\solarradius}}\rp,
\end{equation}
which is much shorter than the full duration of \(\approx\qty{100}{\day} \). Other work arrived at similar estimates for the transient duration \citep[e.g., equation (22) from][]{Yamazaki2017}. %
We conclude that a single mass ejection episode cannot account for the full duration of the \event{} light curve.

\subsection{Powering mechanism}\label{sec:transient:mechanisms}
The hollow points in Figure~\ref{fig:bolometric_properties} show the bolometric properties of \event{} as a function of time with respect to the optical peak \citep[][see Section~\ref{sec:observations:optical_transient} for a summary of how they derived these properties]{De2023}. The light curve has at least two components. First, a plateau in all properties at times \( t\lesssim\qty{10}{\day} \), at the end of which the photosphere cools and shrinks. There is a transition around \( \approx \qty{12}{\day} \) at which the photosphere heats and expands again, reaching a local maximum of \( \approx \qty{4}{\solarradius} \) at \( \approx \qty{25}{\day} \). After that, it gradually contracts at constant affective temperature, with the scatter in the photometric properties increasing drastically after \( \approx \qty{70}{\day} \) as the transient dims.%
Here we examine the decay at times \( > \qty{25}{\day} \), where most of the energy is radiated.

A possible powering mechanism is the recombination of hydrogen in an outflow driven by the energy deposition from the planet. After the planet plunges into the interior, the energy deposition at increasing depths becomes increasingly inefficient at driving an outflow, so the mass loss rate decreases with time. The required mass loss rate to explain the luminosity is approximately
\begin{equation}
\dot{M}_\mathrm{rec}=\frac{m_\mathrm{p}}{X E_\mathrm{H}}L,
\end{equation}
where \( E_\mathrm{H} = \qty{13.6}{\electronvolt} \) is the ionization energy of hydrogen, \( X \approx 0.74 \) is the mass fraction of hydrogen, and \( m_\mathrm{p} \) is the mass of the proton. The dashed black line in Figure~\ref{fig:bolometric_properties} shows the mass loss rate required to explain the light curve entirely via recombination of hydrogen. Since the luminosity of the outflow is proportional to the wind mass loss rate, the implied mass loss rate is \( \dot{M}_\mathrm{rec}\propto L \propto t^{-0.7} \) at late times. The total ejecta mass, assuming hydrogen recombination powers the entire (initial plateau and subsequent decay) of the light curve, is \( \approx \qty{3e-5}{\solarmass} \).

We will now examine a different possible powering mechanism for this part of the light curve. During the surface interaction, the shocks from the planet will impart the ejecta with a distribution of energies. A fraction of the ejecta will remain bound, forming an extended envelope that contracts over the course of its \kh{} time. During this contraction, the envelope radiates the energy that the planet deposited, producing a transient.

Several LRNe show emission consistent with a contracting envelope. After their main outbursts, the luminosity and photosphere radius of V838 Mon and V4332 Sag decreased over several years in a manner consistent with a contracting envelope \citep[see figures 4 and 2, respectively, from][]{Tylenda2005,Tylenda2005a}. In those transients, however, mass ejection on a dynamical timescale powers the light curve on timescales of weeks to months, and the evolution of the material that remains bound to the star dominates the light curve only on timescales of years. Based on a similar comparison to LRNe, \citet{Bear2011a} proposed that a contracting envelope would be responsible for the late-time light curve during a merger between a brown dwarf and a planet. We will now determine whether a contracting envelope can reproduce the \event{} light curve.

We review this model, as presented in \citet{Tylenda2005}, in Appendix~\ref{sec:appendix:envelope}, and discuss its main properties here. We assume that, as a result of the energy deposition, the star forms an extended envelope whose structure can be described as a polytrope with polytropic index \(n=3/2 \). As the envelope contracts, it loses energy. The rate of change of the energy of the envelope---which is related to the rate of change of its radius---is equal in magnitude to its luminosity. This equality results in a differential equation for the envelope radius as a function of time. We assume a constant representative temperature of \qty{5850}{\kelvin} throughout the contraction (shown as a red horizontal dashed line in Figure~\ref{fig:bolometric_properties}). The solid lines in Figure~\ref{fig:bolometric_properties} show a model of a contracting envelope around a sunlike star. The envelope contracts appreciably on its \kh{} timescale,
\begin{equation}
\begin{split}
t_\mathrm{KH}=&\frac{G \menv \mstar}{2\renv L_\mathrm{env}}\\
=&\qty{40}{\day}\lp\frac{\menv}{\qty{e-6}{\solarmass}}\rp
\lp\frac{\mstar}{\unit{\solarmass}}\rp
\lp\frac{\renv}{\qty{5}{\solarradius}}\rp^{-1}\\
&\times\lp\frac{L_\mathrm{env}}{\qty{e35}{\erg\per\second}}\rp^{-1},
\end{split}
\end{equation}
We set the initial radius and luminosity of the envelope to the observed values and choose the mass of the envelope such that the contraction timescale of the envelope matches the observations. That envelope mass is \(\menv = \qty{1.1e-6}{\solarmass} \). We also show for comparison envelopes three times less and more massive. The contracting envelope model can reasonably reproduce the late-time \(\propto t^{-0.7} \) decay of the light curve.

In reality, both of these powering mechanisms likely play a role, with some of the shocked material becoming unbound and forming a recombination wind, and some of the material remaining bound and forming a contracting envelope.

\subsection{Future evolution}
On timescales longer than the initial transient, any energy deposited deeper in the star becomes the dominant perturbation to the intrinsic stellar luminosity \citep{Metzger2012}. The deeper the energy deposition, the longer it will take to reach the surface. The star will return to its original state only on the much longer \kh{} time of the deepest heated region. Therefore, follow-up observations could constrain the properties of the heating. For example, if the star returns to its pre-merger luminosity on timescales much shorter than its global \kh{} time, it would rule out a large thermal perturbation deep in the star.

The exact association between the timescale over which the star returns to its unperturbed state and the properties of the planet is less straightforward. In principle, more resilient planets could heat deeper regions of the star (although the details depend on the dynamical and energetic distribution of the planetary debris). Although a less massive but denser planet will survive deeper in the star, the luminosity perturbation from the cooling of the deeper layers might be so small---because the planet is small---that the perturbed luminosity becomes observationally indistinguishable from the unperturbed one much sooner than the \kh{} time of the deepest heated region\@. \citet{Metzger2017} considered a star-planet merger as a potential explanation for the secular dimming of KIC 8462852 \citep{Boyajian2016}, an F-type main-sequence star. They used a one-dimensional stellar evolution code to model the evolution of the star following heating by a merger with companions ranging from Io to a \qty{50}{\jupitermass} brown dwarf. The star took roughly a hundred times longer to return to its original luminosity if the companion was a brown dwarf rather than a Jupiter. However, even though the Earth survived deeper than the Jupiter because it is denser (they considered only destruction via tidal disruption), its orbital energy is so much smaller that the luminosity of the star became similar to the unperturbed luminosity sooner than for the Jupiter \citep[see Figures~1~and~2 from][]{Metzger2017}.

During the merger, the planet also deposits its angular momentum into the star. Therefore, as a result of angular momentum conservation after the merger, another prediction of the contracting envelope model is that the surface rotation rate of the star should increase as \( R_\mathrm{phot} ^ {-2} \). However, the dusty obscuration of the photosphere in the \citet{De2023} spectra makes the rotation rate shortly after the merger uncertain.

\subsection{Alternative scenarios}
\subsubsection{Tidal disruption event}\label{sec:alt:tde}
Another possible outcome of the star-planet interaction is the tidal disruption of the planet above the stellar surface\@. \citet{Bear2011} and \citet{Metzger2012} studied the transients arising from the tidal disruption of a planet by a brown dwarf and a star, respectively. Qualitatively, the evolution is as follows \citep{Metzger2012}: the debris of the disrupted planet forms a disk around the star. The accretion rate is higher than the Eddington accretion rate, resulting in an outflow with a characteristic luminosity
\begin{equation}
L \approx \qty{e37}{\erg\per\second}\lp\mplanet/\unit{\jupitermass}\rp.\label{eq:l_super_eddington_wind}
\end{equation}
Once the accretion rate is below the Eddington rate, which occurs after a time \( \approx t_\mathrm{Edd}\approx\qty{80}{\day}\lp\mplanet/\unit{\jupitermass}\rp^{3/4}\), the emission from the accretion disk becomes directly visible. At that point, the transient becomes brighter and hotter, since the effective temperature in the disk at radial coordinate \( r \) is \citep[see equation~(29) from][]{Metzger2012}
\begin{equation}
T_{\rm acc} \approx \qty{6.6e4}{\kelvin}\lp\frac{\dot{M}}{\dot{M}_\mathrm{Edd}}\rp^{1/4}\lp\frac{r}{\unit{\solarradius}}\rp^{-3/4}.
\end{equation}
In \event{}, however, the effective temperature remains \( \lesssim \qty{e4}{\kelvin} \) throughout the transient \citep[Figure~\ref{fig:bolometric_properties} and][]{De2023}, suggesting that the emission is not arising from a hot accretion disk. If we instead let \( t_\mathrm{Edd} > \qty{100}{\day} \) (by setting \( \mplanet \gtrsim \unit{\jupitermass} \)), so that the observed transient is a result of the super-Eddington wind (before the disk becomes visible), the luminosity of that wind (equation~\eqref{eq:l_super_eddington_wind}, \(\gtrsim\qty{e37}{\erg\per\second}\)) would be much larger than the observed one (\(\approx\qty{e35}{\erg\per\second}\)). We therefore consider it unlikely that \event{} is a planetary tidal disruption event.

\subsubsection{Jet formation}
Jets are a potential powering mechanism for merger transients \citep[e.g.,][]{Soker2020,Soker2021a}\@. \citet{Soker2023} studied the potential role on jets in \event{}, and suggested that the planet could have accreted material from the star during the pre-merger epochs and launched a jet. To allow accretion, the relative velocity between the stellar surface and the planet must be smaller than we have assumed; in particular, it must be much smaller than the orbital velocity, so that gravitational capture can form an accretion disk around the planet.
In this scenario, the terminal velocity of jet ejecta is approximately the escape speed from the object launching the jet. Since, for a planet \(\approx\qty{10}{\jupitermass}\), the escape velocity of the planet and of the star are of the same order, some jet material might become unbound but have a small terminal speed\@. \citet{Soker2023} suggested that this process could explain the \(\approx\qty{35}{\kilo\meter\per\second}\) expansion speed of the inner edge of the dust shell \citep[Section~\ref{sec:observations:optical_transient} and][]{De2023}. A prediction from this scenario is that \event{} will form a bipolar nebula.

\subsubsection{Accretion outburst onto a young stellar object}
\citet{De2023} considered the possibility that \event{} was the result of an accretion episode onto a young stellar object \citep[YSO; for a review on accretion onto young stars, see][]{Hartmann2016}. They found it unlikely because the optical and near-IR spectra lacked the atomic emission lines characteristic of hot accreting gas. The estimated evolutionary stage of the \event{} star (on or slightly beyond the main sequence) sets it apart from two other star-planet merger candidates: ASASSN-15qi \citep{Herczeg2016,Kashi2017} and ASASSN-13db \citep{SiciliaAguilar2017,Kashi2018,Kashi2019}. The stars in those transients are YSOs, and they experienced other outbursts within decades of the potential star-planet outburst. The possibility of accretion outbursts in those two sources makes the unambiguous determination of the cause of a particular outburst more challenging.

\section{Energetic constraints}\label{sec:energetics}
We will now combine the results of previous sections with the observed energetic properties of \event{}. The orbital energy of the planet powers the transient, so the energetics can constrain the mass of the planet. We will consider the constraints from both the pre-merger observations and from the light curve during the main transient.

\subsection{From the pre-merger observations}
There are observational constraints of the ejecta mass at three epochs. We summarized those constraints in Section~\ref{sec:observations}.
The energy required to produce ejecta is
\begin{equation}
\begin{split}
E_\mathrm{ej}=\frac{1}{2}M_\mathrm{ej}\vesc^2\approx&\qty{3.79e44}{\erg}\lp\frac{M_\mathrm{ej}}{\qty{e-4}{\solarmass}}\rp \\
&\lp\frac{\mstar}{\unit{\solarmass}}\rp
\lp\frac{\rstar}{\unit{\solarradius}}\rp^{-1},
\end{split}
\end{equation}

In Section~\ref{sec:merger:ejecta} we studied the scenario in which the pre-merger ejecta is produced as the planet interacts with the stellar surface. Figure~\ref{fig:shocked_mass} shows the mass that the planet has shocked as a function of epoch, for planets of different masses. Given the small fraction of the orbital energy that is available in the small change in orbital separation between the two epochs, this mechanism requires a massive planet. In that section, we estimated that planets with masses
\begin{equation}
\mplanet\gtrsim\qty{5}{\jupitermass}\label{eq:m_planet_premerger}
\end{equation}
are necessary to meet the pre-merger ejecta constraints\@. \citet{De2023} arrived at a similar estimate \( \mplanet \approx \qty{10}{\jupitermass} \) by extrapolating models of pre-merger mass loss in stellar mergers \citep{MacLeod2020}.

\subsection{From the main outburst}
The radiated energy during the transient is \( \qty{6.5e41}{\erg} \). Constraining the orbital energy from the radiated energy requires the radiative efficiency, which can vary by orders of magnitude depending on the process responsible for the radiation. For example, consider the recombination wind and the contracting envelope we discussed in Section~\ref{sec:transient}. The energy radiated by the contracting envelope is equal to the energy it took to inflate it, so the radiative efficiency is unity. In contrast, to form a recombination wind, the gas in the wind must be effectively unbound. The energy released by recombination is only a fraction
\begin{equation}
\varepsilon_\mathrm{rec}=\frac{\qty{13.6}{\electronvolt}}{m_\mathrm{p}\vesc^2}\approx3\times10^{-3}\label{eq:recombination_efficiency}
\end{equation}
of the energy required to unbind the gas, so the radiative efficiency is smaller.

If the entire light curve is the result of a recombination wind, the implied ejecta mass is
\begin{equation}
M_\mathrm{ej}=\frac{E_\mathrm{rad}}{E_\mathrm{H}}\frac{m_\mathrm{p}}{X}\approx\qty{3.4e-5}{\solarmass}\label{eq:m_ej_wind}.
\end{equation}
From equation~\ref{eq:minimum_planet_mass_ejecta} and the planet mass-radius relation the planet mass that corresponds to this amount of ejecta is
\begin{gather}
\mplanet\approx\qty{0.3}{\jupitermass},\label{eq:m_planet_rec}\\
\rplanet\approx\qty{1}{\jupiterradius}.
\end{gather}
In contrast, if we assume a radiative efficiency of unity, the constraint on the planet mass follows from
\begin{equation}
\left.\frac{d\eorb}{da}\right|_{a=\rstar}z = E_\mathrm{rad},\\
\end{equation}
yielding
\begin{gather}
\mplanet\approx\qty{3.5}{\earthmass},\label{eq:m_planet_epsrad_1}\\
\rplanet\approx\qty{1.7}{\earthradius},
\end{gather}
where we set \( z = 2\rplanet \).

The planet mass constraints in equations~\eqref{eq:m_planet_rec}~and~\eqref{eq:m_planet_epsrad_1} are smaller than the pre-merger constraint from equation~\eqref{eq:m_planet_premerger}. This difference warrants further examination, as it is unclear whether the larger planetary masses we estimate from the pre-merger observations are consistent with the merger light curve. Figure~\ref{fig:shocked_mass} shows that the rate at which the planet shocks stellar material increases steeply as the orbital decay time decreases. A significant fraction of the total ejecta is likely produced in the last few orbital periods before the planet plunges into the stellar interior \citep{Metzger2012,Lau2025}. The recombination transient associated with this mass ejection on dynamical timescales has a characteristic duration of hours \citep{Yamazaki2017}, which is shorter than the ZTF cadence of \( \approx \) days. It is therefore possible that a fraction of the ejecta of the transient was produced over a timescale inaccessible to ZTF\@.

While some dynamical ejecta might not appear in the transient light curve, it should eventually form dust, for which there are post-merger observational constraints. These dust constraints arise from the evolution of the SED, not from the light curve, so they are free from cadence effects\@. As we discussed in Section~\ref{sec:observations:optical_transient}, \citet{De2023} and \citet{Lau2025a} estimated the ejecta mass to be \( \approx \qty{e-4}{\solarmass} \) (see equation~\eqref{eq:m_ej_obs_320}). This amount is slightly larger than required by a recombination wind (the least radiatively efficient mechanism we consider, equation~\eqref{eq:m_ej_wind}), supporting the idea that the light curve does not capture at least some dynamical mass ejection. The observed ejecta mass suggests that the electromagnetic signatures of only a fraction of the ejecta appear in the ZTF light curve, ranging from \( \approx \) a percent (in the cooling-envelope model) to \( \approx \) tens of percents (in the recombination wind model).

Equation~\eqref{eq:minimum_planet_mass_ejecta} shows that a planet at least as massive as Jupiter is needed to produce the observed \( \approx \qty{e-4}{\solarmass} \) of post-merger ejecta. The uncertainties in the lower end of the observed ejecta mass are significant, but the best-fit value of \( \approx \qty{e-4}{\solarmass} \) energetically rules out a planet with mass between Earth and Neptune (e.g., equation~\ref{eq:m_planet_epsrad_1}). For this reason, we favor a planet at least as massive as Jupiter.

\section{Conclusions}\label{sec:summary}
\citet{De2023} interpreted the \event{} transient as a star-planet merger. Here, we explored this possibility in more detail through models of the system before and during the merger. We used the pre-merger dust formation observations to estimate the mass of the planet to be \( \gtrsim \qty{5}{\jupitermass} \). We argued that the most promising mechanisms responsible for the light curve are the contraction of an inflated envelope around the merger remnant or the recombination of hydrogen in an outflow. We also argued that some of the ejecta was produced on a dynamical timescale and is unobservable in the ZTF light curve, but observable through post-merger dust formation. We summarize our results in more detail below.

The pre-merger evolution likely consists of a planet whose orbit decayed as a result of tidal dissipation. The archival image and the small photosphere radius during the transient support the idea that planets can merge with their host stars during the main sequence or early during the post-main-sequence. At the time of the archival image, the tidal interactions between the star and the planet had not affected the appearance of the star (Figure~\ref{fig:tides_observability}).

Once the star and the planet come into contact, drag forces affect the orbital decay of the planet. The planet shocks the stellar material at the surface, ejecting some of it. This mechanism can account for the pre-merger ejecta if the mass of the planet is \( \gtrsim \qty{5}{\jupitermass} \) (Figure~\ref{fig:shocked_mass}). A planet in this mass range has a mean density of \(\gtrsim\qty{5}{\jupitermass}\), much higher than the \( \approx\qty{1}{\gram\per\cubic\centi\meter}\) of the roughly sunlike star in \event{}. This density contrast is consistent with the planet avoiding tidal disruption above the stellar surface. Energy deposited deep in the stellar interior, if any, likely reaches the surface only on timescales longer than the duration of the transient.

The duration of the light curve suggests that \event{} cannot be powered by a single episode of mass ejection on a dynamical timescale (Figure~\ref{fig:transient_timescales}). It is possible, however, that the planet ejected mass dynamically as it plunged into the stellar interior, as predicted by previous work \citep[e.g.,][]{Metzger2012,Yamazaki2017,Lau2025}. The timescale of this dynamical transient \( \approx \) hours might be shorter than the cadence of the observations. The light curve can be the result of a recombination wind with a decreasing mass loss rate (requiring \( \qty{3.4e-5}{\solarmass} \) of outflow), or as the contraction of a remnant inflated envelope of \( \approx \qty{e-6}{\solarmass} \). Likely, both mechanisms play a role, with some shocked material becoming unbound and recombining, and some remaining bound and contracting gradually. The detection of circumstellar gas around the remnant \citep{Lau2025} indeed suggests that some shocked material remains bound.

Estimates of the ejecta mass at the \qty{320}{\day} epoch of \( \approx \qty{e-4}{\solarmass} \)---larger than implied by a recombination wind---tentatively support the idea that some dynamical mass loss is absent from the light curve. Energetically, the best-fit value for the mass of the observed post-merger ejecta suggests a planet at least as massive as Jupiter. Combined with the pre-merger ejecta constraints, the observed ejecta masses suggest \event{} was the result of a merger between a star and a planet with mass at least several times that of Jupiter. Future models combining hydrodynamics and radiative transfer could improve our understanding of these mergers and determine the properties of their progenitor systems.

\begin{acknowledgments}
We thank Elisabeth~Adams, Andrea~Antoni, Evan~Bauer, Matteo~Cantiello, Rosa~Wallace~Everson, Brian~Jackson, Seth~Jacobson, Ryan~Lau, Abraham~Loeb, Brian~Metzger, Sarai~Rankin, Melinda~Soares-Furtado, Alexander~Stephan, and Ashley~Villar for discussions. We thank Kishalay~De for discussions and for sharing the observation data in Figure~\ref{fig:bolometric_properties}.
RY is grateful for support from a Doctoral Fellowship from the University of California Institute for Mexico and the United States (UCMEXUS), a Texas Advanced Computing Center (TACC) Frontera Computational Science Fellowship, and a NASA FINESST award. This research was possible thanks to funding at UC Santa Cruz through the Heising--Simons Foundation, NSF grants: AST 1852393, AST 2150255 and AST 2206243. %
M.M. gratefully acknowledges support from the Clay Postdoctoral Fellowship of the Smithsonian Astrophysical Observatory. B.I. acknowledges support from the University of California President's Postdoctoral Fellowship (PPFP) and the Vera Rubin Postdoctoral Fellowship. This research has made use of NASA's Astrophysics Data System Bibliographic Services.
\end{acknowledgments}

\software{Mathematica, matplotlib \citep{Hunter2007}, MESA r24.08.1 \citep{Fuller1985,Iglesias1993,Oda1994,Saumon1995,Iglesias1996,Itoh1996,Angulo1999,Langanke2000,Timmes2000,Rogers2002,Irwin2004,Ferguson2005,Cassisi2007,Chugunov2007,Cyburt2010,Potekhin2010,Paxton2011,Paxton2013,Paxton2015,Poutanen2017,Paxton2018,Paxton2019,Blouin2020,Jermyn2021,Jermyn2023}, MESA SDK 24.7.1 \citep{Townsend2024}, numpy \citep{Harris2020}, pandas \citep{McKinney2010}, py\_mesa\_reader \citep{Schwab2024}, scipy \citep{Virtanen2020}, unyt \citep{Goldbaum2018}.}

\appendix
\section{Ram pressure drag}\label{sec:appendix:drag}
We approximate the drag force as the integral the momentum flux over the \crosssection{} of the planet,
\begin{equation}
F = \int \rho \vorb^2 d\sigma.
\end{equation}
Approximating the orbital speed as constant across the planet, and the density profile as exponential with a scale height at the substellar point \hrho{}, we obtain
\begin{equation}
\label{eq:drag}
F_\mathrm{d} =\rho\vorb^2\sigma,
\end{equation}
where \(\rho \) is the density at the substellar point of the planet, and
\begin{equation}
\sigma\equiv2\pi I_1\lp\epsrho\rp e^{-\epsrho}\hrho\rplanet
\label{eq:effective_cross_section}
\end{equation}
is the effective \crosssection{} of the planet, where \(I_1 \) is the modified Bessel function of the first kind. If the flow is heterogeneous on scales smaller than the size of the planet (\(\epsrho\gg1 \)), then \(I_1\lp\epsrho\rp \simeq \exp\lp \epsrho\rp/\sqrt{2\pi \epsrho} \), and the effective \crosssection{} of the planet becomes \citep[see also][]{Metzger2012}
\begin{equation}
\sigma \simeq \sqrt{2\pi}\hrho^{3/2}\rplanet^{1/2} = \sqrt{2\pi}\epsrho^{-3/2}\rplanet^2\label{eq:sigma_high_epsrho}.
\end{equation}
In contrast, when the density is approximately constant across the planet (\(\epsrho\ll1 \)), then \(I_1\lp\epsrho\rp\approx\epsrho/2 \), and the effective \crosssection{} approaches the typical value
\begin{equation}
\sigma \approx \pi\rplanet^2.
\end{equation}
Figure~\ref{fig:dimensionless_quantities} shows that during the surface interaction, \( \epsrho > 1\), such that equation~\eqref{eq:sigma_high_epsrho} is the effective \crosssection{} of the planet.

\section{Cooling envelope model}\label{sec:appendix:envelope}
\citet{Tylenda2005a} derived a differential equation for the evolution of a contracting polytropic envelope. See their Appendix A for a derivation\@. \citet{Tylenda2005a} showed that the differential equation for the radius of the envelope is (their equations (A.18 and A.19))
\begin{equation}
\frac{8\pi\renv^3\sigma_\mathrm{SB}\teff^4}{G\menv\mstar}=\left[\frac{d}{dt}\lp\frac{I_\mathrm{e}}{I_\mathrm{m}}\rp-\frac{\dot{R}_\mathrm{env}}{\renv}\frac{\ieint}{\imint}\right].\label{eq:ode_tylenda}
\end{equation}
Here, \renv{} is the radius of the envelope, \( \sigma_\mathrm{SB} \) is the \stefanboltzmann{} constant, \teff{} is the effective temperature of the envelope, \menv{} is the mass of the envelope, and
\begin{gather}
\ieint\equiv\int_{x_\star}^{1}\lp1-x\rp^n x^{1-n}\,dx,\\
\imint\equiv\int_{x_\star}^{1}\lp1-x\rp^n x^{2-n}\,dx,
\end{gather}
where \( x_\star \equiv \rstar/\renv \). We can simplify equation~\eqref{eq:ode_tylenda} by using
\begin{equation}
\frac{d}{d\renv}\int_{x_\star}^{1}f\lp x\rp \, dx=\frac{x_\star}{\renv}f\lp x_\star\rp,
\end{equation}
from which we obtain
\begin{gather}
\frac{d\ieint}{d\renv}=\renv^{-1}\lp 1 - x_\star \rp^n x_\star^{2-n},\\
\frac{d\imint}{d\renv}=\renv^{-1}\lp 1 - x_\star \rp^n x_\star^{3-n}.
\end{gather}
Using these relations, we can rewrite equation~\eqref{eq:ode_tylenda} as
\begin{equation}
\dot{R}_\mathrm{env}=-\frac{\renv}{t_\mathrm{KH,env}}\frac{1}{g\lp x_\star\rp},
\end{equation}
where
\begin{equation}
\begin{split}
g\lp x_\star\rp\equiv&
\frac{\ieint}{\imint}
+\frac{\ieint}{\imint^2}\lp 1- x_\star \rp^n x_\star^{3-n}\\
&-\frac{\lp 1 - x_\star \rp^n x_\star^{2-n}}{\imint}
\end{split}
\end{equation}
and
\begin{equation}
t_\mathrm{KH,env}\equiv\frac{G\menv\mstar}{8\pi\renv^3\sigma_\mathrm{SB}\teff^4}.
\end{equation}

\bibliographystyle{aasjournal}
\bibliography{bib}

\begin{thebibliography}{}
\expandafter\ifx\csname natexlab\endcsname\relax\def\natexlab#1{#1}\fi
\providecommand{\url}[1]{\href{#1}{#1}}
\providecommand{\dodoi}[1]{doi:~\href{http://doi.org/#1}{\nolinkurl{#1}}}
\providecommand{\doeprint}[1]{\href{http://ascl.net/#1}{\nolinkurl{http://ascl.net/#1}}}
\providecommand{\doarXiv}[1]{\href{https://arxiv.org/abs/#1}{\nolinkurl{https://arxiv.org/abs/#1}}}

\bibitem[{M. {Adam{\'o}w} {et~al.}(2012){Adam{\'o}w}, {Niedzielski}, {Villaver}, {Nowak}, \& {Wolszczan}}]{Adamow2012}
{Adam{\'o}w}, M., {Niedzielski}, A., {Villaver}, E., {Nowak}, G., \& {Wolszczan}, A. 2012, \bibinfo{title}{{BD+48 740{\textemdash}Li Overabundant Giant Star with a Planet: A Case of Recent Engulfment?},} \apjl, 754, L15, \dodoi{10.1088/2041-8205/754/1/L15}

\bibitem[{C. {Aguilera-G{\'o}mez} {et~al.}(2020){Aguilera-G{\'o}mez}, {Chanam{\'e}}, \& {Pinsonneault}}]{AguileraGomez2020}
{Aguilera-G{\'o}mez}, C., {Chanam{\'e}}, J., \& {Pinsonneault}, M.~H. 2020, \bibinfo{title}{{On Lithium-6 as a Diagnostic of the Lithium-enrichment Mechanism in Red Giants},} \apjl, 897, L20, \dodoi{10.3847/2041-8213/ab9d26}

\bibitem[{C. {Aguilera-G{\'o}mez} {et~al.}(2016{\natexlab{a}}){Aguilera-G{\'o}mez}, {Chanam{\'e}}, {Pinsonneault}, \& {Carlberg}}]{AguileraGomez2016}
{Aguilera-G{\'o}mez}, C., {Chanam{\'e}}, J., {Pinsonneault}, M.~H., \& {Carlberg}, J.~K. 2016{\natexlab{a}}, \bibinfo{title}{{On Lithium-rich Red Giants: Engulfment on the Giant Branch of Trumpler 20},} \apjl, 833, L24, \dodoi{10.3847/2041-8213/833/2/L24}

\bibitem[{C. {Aguilera-G{\'o}mez} {et~al.}(2016{\natexlab{b}}){Aguilera-G{\'o}mez}, {Chanam{\'e}}, {Pinsonneault}, \& {Carlberg}}]{AguileraGomez2016a}
{Aguilera-G{\'o}mez}, C., {Chanam{\'e}}, J., {Pinsonneault}, M.~H., \& {Carlberg}, J.~K. 2016{\natexlab{b}}, \bibinfo{title}{{On Lithium-rich Red Giants. I. Engulfment of Substellar Companions},} \apj, 829, 127, \dodoi{10.3847/0004-637X/829/2/127}

\bibitem[{J.~B. {Alexander}(1967){Alexander}}]{Alexander1967}
{Alexander}, J.~B. 1967, \bibinfo{title}{{A possible source of lithium in the atmospheres of some red giants},} The Observatory, 87, 238

\bibitem[{C. {Angulo} {et~al.}(1999){Angulo}, {Arnould}, {Rayet}, {Descouvemont}, {Baye}, {Leclercq-Willain}, {Coc}, {Barhoumi}, {Aguer}, {Rolfs}, {Kunz}, {Hammer}, {Mayer}, {Paradellis}, {Kossionides}, {Chronidou}, {Spyrou}, {degl'Innocenti}, {Fiorentini}, {Ricci}, {Zavatarelli}, {Providencia}, {Wolters}, {Soares}, {Grama}, {Rahighi}, {Shotter}, \& {Lamehi Rachti}}]{Angulo1999}
{Angulo}, C., {Arnould}, M., {Rayet}, M., {et~al.} 1999, \bibinfo{title}{{A compilation of charged-particle induced thermonuclear reaction rates},} \nphysa, 656, 3, \dodoi{10.1016/S0375-9474(99)00030-5}

\bibitem[{A. {Bailey} \& J. {Goodman}(2019){Bailey} \& {Goodman}}]{Bailey2019}
{Bailey}, A., \& {Goodman}, J. 2019, \bibinfo{title}{{Understanding WASP-12b},} \mnras, 482, 1872, \dodoi{10.1093/mnras/sty2805}

\bibitem[{A.~J. {Barker}(2020){Barker}}]{Barker2020}
{Barker}, A.~J. 2020, \bibinfo{title}{{Tidal dissipation in evolving low-mass and solar-type stars with predictions for planetary orbital decay},} \mnras, 498, 2270, \dodoi{10.1093/mnras/staa2405}

\bibitem[{A.~J. {Barker} {et~al.}(2024){Barker}, {Efroimsky}, {Makarov}, \& {Veras}}]{Barker2024}
{Barker}, A.~J., {Efroimsky}, M., {Makarov}, V.~V., \& {Veras}, D. 2024, \bibinfo{title}{{On the orbital decay of the gas giant Kepler-1658b},} \mnras, 527, 5131, \dodoi{10.1093/mnras/stad3530}

\bibitem[{E. {Bear} {et~al.}(2011){Bear}, {Kashi}, \& {Soker}}]{Bear2011a}
{Bear}, E., {Kashi}, A., \& {Soker}, N. 2011, \bibinfo{title}{{Mergerburst transients of brown dwarfs with exoplanets},} \mnras, 416, 1965, \dodoi{10.1111/j.1365-2966.2011.19171.x}

\bibitem[{E. {Bear} \& N. {Soker}(2011){Bear} \& {Soker}}]{Bear2011}
{Bear}, E., \& {Soker}, N. 2011, \bibinfo{title}{{Evaporation of Jupiter-like planets orbiting extreme horizontal branch stars},} \mnras, 414, 1788, \dodoi{10.1111/j.1365-2966.2011.18527.x}

\bibitem[{A. {Behmard} {et~al.}(2023){Behmard}, {Dai}, {Brewer}, {Berger}, \& {Howard}}]{Behmard2023}
{Behmard}, A., {Dai}, F., {Brewer}, J.~M., {Berger}, T.~A., \& {Howard}, A.~W. 2023, \bibinfo{title}{{Planet engulfment detections are rare according to observations and stellar modelling},} \mnras, 521, 2969, \dodoi{10.1093/mnras/stad745}

\bibitem[{S. {Blouin} {et~al.}(2020){Blouin}, {Shaffer}, {Saumon}, \& {Starrett}}]{Blouin2020}
{Blouin}, S., {Shaffer}, N.~R., {Saumon}, D., \& {Starrett}, C.~E. 2020, \bibinfo{title}{{New Conductive Opacities for White Dwarf Envelopes},} \apj, 899, 46, \dodoi{10.3847/1538-4357/ab9e75}

\bibitem[{T.~S. {Boyajian} {et~al.}(2016){Boyajian}, {LaCourse}, {Rappaport}, {Fabrycky}, {Fischer}, {Gandolfi}, {Kennedy}, {Korhonen}, {Liu}, {Moor}, {Olah}, {Vida}, {Wyatt}, {Best}, {Brewer}, {Ciesla}, {Cs{\'a}k}, {Deeg}, {Dupuy}, {Handler}, {Heng}, {Howell}, {Ishikawa}, {Kov{\'a}cs}, {Kozakis}, {Kriskovics}, {Lehtinen}, {Lintott}, {Lynn}, {Nespral}, {Nikbakhsh}, {Schawinski}, {Schmitt}, {Smith}, {Szabo}, {Szabo}, {Viuho}, {Wang}, {Weiksnar}, {Bosch}, {Connors}, {Goodman}, {Green}, {Hoekstra}, {Jebson}, {Jek}, {Omohundro}, {Schwengeler}, \& {Szewczyk}}]{Boyajian2016}
{Boyajian}, T.~S., {LaCourse}, D.~M., {Rappaport}, S.~A., {et~al.} 2016, \bibinfo{title}{{Planet Hunters IX. KIC 8462852 - where's the flux?},} \mnras, 457, 3988, \dodoi{10.1093/mnras/stw218}

\bibitem[{R.~M. {Cabez{\'o}n} {et~al.}(2023){Cabez{\'o}n}, {Abia}, {Dom{\'\i}nguez}, \& {Garc{\'\i}a-Senz}}]{Cabezon2023}
{Cabez{\'o}n}, R.~M., {Abia}, C., {Dom{\'\i}nguez}, I., \& {Garc{\'\i}a-Senz}, D. 2023, \bibinfo{title}{{Sub-stellar engulfment by a main-sequence star: Where is the lithium?},} \aap, 670, A155, \dodoi{10.1051/0004-6361/202244848}

\bibitem[{A.~G.~W. {Cameron} \& W.~A. {Fowler}(1971){Cameron} \& {Fowler}}]{Cameron1971}
{Cameron}, A.~G.~W., \& {Fowler}, W.~A. 1971, \bibinfo{title}{{Lithium and the s-PROCESS in Red-Giant Stars},} \apj, 164, 111, \dodoi{10.1086/150821}

\bibitem[{J.~K. {Carlberg} {et~al.}(2012){Carlberg}, {Cunha}, {Smith}, \& {Majewski}}]{Carlberg2012}
{Carlberg}, J.~K., {Cunha}, K., {Smith}, V.~V., \& {Majewski}, S.~R. 2012, \bibinfo{title}{{Observable Signatures of Planet Accretion in Red Giant Stars. I. Rapid Rotation and Light Element Replenishment},} \apj, 757, 109, \dodoi{10.1088/0004-637X/757/2/109}

\bibitem[{J.~K. {Carlberg} {et~al.}(2013){Carlberg}, {Cunha}, {Smith}, \& {Majewski}}]{Carlberg2013}
{Carlberg}, J.~K., {Cunha}, K., {Smith}, V.~V., \& {Majewski}, S.~R. 2013, \bibinfo{title}{{Li-enrichment in red giant rapid rotators: Planet engulfment versus extra mixing},} Astronomische Nachrichten, 334, 120, \dodoi{10.1002/asna.201211757}

\bibitem[{J.~K. {Carlberg} {et~al.}(2009){Carlberg}, {Majewski}, \& {Arras}}]{Carlberg2009}
{Carlberg}, J.~K., {Majewski}, S.~R., \& {Arras}, P. 2009, \bibinfo{title}{{The Role of Planet Accretion in Creating the Next Generation of Red Giant Rapid Rotators},} \apj, 700, 832, \dodoi{10.1088/0004-637X/700/1/832}

\bibitem[{J.~K. {Carlberg} {et~al.}(2011){Carlberg}, {Majewski}, {Arras}, {Smith}, {Cunha}, \& {Bizyaev}}]{Carlberg2011a}
{Carlberg}, J.~K., {Majewski}, S.~R., {Arras}, P., {et~al.} 2011, in American Institute of Physics Conference Series, Vol. 1331, Planetary Systems Beyond the Main Sequence, ed. S.~{Schuh}, H.~{Drechsel}, \& U.~{Heber} (AIP), 33--40, \dodoi{10.1063/1.3556182}

\bibitem[{J.~K. {Carlberg} {et~al.}(2010){Carlberg}, {Smith}, {Cunha}, {Majewski}, \& {Rood}}]{Carlberg2010}
{Carlberg}, J.~K., {Smith}, V.~V., {Cunha}, K., {Majewski}, S.~R., \& {Rood}, R.~T. 2010, \bibinfo{title}{{The Super Lithium-rich Red Giant Rapid Rotator G0928+73.2600: A Case for Planet Accretion?},} \apjl, 723, L103, \dodoi{10.1088/2041-8205/723/1/L103}

\bibitem[{A.~R. {Casey} {et~al.}(2019){Casey}, {Ho}, {Ness}, {Hogg}, {Rix}, {Angelou}, {Hekker}, {Tout}, {Lattanzio}, {Karakas}, {Woods}, {Price-Whelan}, \& {Schlaufman}}]{Casey2019}
{Casey}, A.~R., {Ho}, A. Y.~Q., {Ness}, M., {et~al.} 2019, \bibinfo{title}{{Tidal Interactions between Binary Stars Can Drive Lithium Production in Low-mass Red Giants},} \apj, 880, 125, \dodoi{10.3847/1538-4357/ab27bf}

\bibitem[{S. {Cassisi} {et~al.}(2007){Cassisi}, {Potekhin}, {Pietrinferni}, {Catelan}, \& {Salaris}}]{Cassisi2007}
{Cassisi}, S., {Potekhin}, A.~Y., {Pietrinferni}, A., {Catelan}, M., \& {Salaris}, M. 2007, \bibinfo{title}{{Updated Electron-Conduction Opacities: The Impact on Low-Mass Stellar Models},} \apj, 661, 1094, \dodoi{10.1086/516819}

\bibitem[{K.~C. {Chambers} {et~al.}(2016){Chambers}, {Magnier}, {Metcalfe}, {Flewelling}, {Huber}, {Waters}, {Denneau}, {Draper}, {Farrow}, {Finkbeiner}, {Holmberg}, {Koppenhoefer}, {Price}, {Rest}, {Saglia}, {Schlafly}, {Smartt}, {Sweeney}, {Wainscoat}, {Burgett}, {Chastel}, {Grav}, {Heasley}, {Hodapp}, {Jedicke}, {Kaiser}, {Kudritzki}, {Luppino}, {Lupton}, {Monet}, {Morgan}, {Onaka}, {Shiao}, {Stubbs}, {Tonry}, {White}, {Ba{\~n}ados}, {Bell}, {Bender}, {Bernard}, {Boegner}, {Boffi}, {Botticella}, {Calamida}, {Casertano}, {Chen}, {Chen}, {Cole}, {Deacon}, {Frenk}, {Fitzsimmons}, {Gezari}, {Gibbs}, {Goessl}, {Goggia}, {Gourgue}, {Goldman}, {Grant}, {Grebel}, {Hambly}, {Hasinger}, {Heavens}, {Heckman}, {Henderson}, {Henning}, {Holman}, {Hopp}, {Ip}, {Isani}, {Jackson}, {Keyes}, {Koekemoer}, {Kotak}, {Le}, {Liska}, {Long}, {Lucey}, {Liu}, {Martin}, {Masci}, {McLean}, {Mindel}, {Misra}, {Morganson}, {Murphy}, {Obaika}, {Narayan}, {Nieto-Santisteban}, {Norberg}, {Peacock}, {Pier}, {Postman}, {Primak}, {Rae}, {Rai}, {Riess}, {Riffeser}, {Rix}, {R{\"o}ser}, {Russel}, {Rutz}, {Schilbach}, {Schultz}, {Scolnic}, {Strolger}, {Szalay}, {Seitz}, {Small}, {Smith}, {Soderblom}, {Taylor}, {Thomson}, {Taylor}, {Thakar}, {Thiel}, {Thilker}, {Unger}, {Urata}, {Valenti}, {Wagner}, {Walder}, {Walter}, {Watters}, {Werner}, {Wood-Vasey}, \& {Wyse}}]{Chambers2016}
{Chambers}, K.~C., {Magnier}, E.~A., {Metcalfe}, N., {et~al.} 2016, \bibinfo{title}{{The Pan-STARRS1 Surveys},} arXiv e-prints, arXiv:1612.05560, \dodoi{10.48550/arXiv.1612.05560}

\bibitem[{A.~I. {Chugunov} {et~al.}(2007){Chugunov}, {Dewitt}, \& {Yakovlev}}]{Chugunov2007}
{Chugunov}, A.~I., {Dewitt}, H.~E., \& {Yakovlev}, D.~G. 2007, \bibinfo{title}{{Coulomb tunneling for fusion reactions in dense matter: Path integral MonteCarlo versus mean field},} \prd, 76, 025028, \dodoi{10.1103/PhysRevD.76.025028}

\bibitem[{R.~P. {Church} {et~al.}(2020){Church}, {Mustill}, \& {Liu}}]{Church2020}
{Church}, R.~P., {Mustill}, A.~J., \& {Liu}, F. 2020, \bibinfo{title}{{Super-Earth ingestion can explain the anomalously high metal abundances of M67 Y2235},} \mnras, 491, 2391, \dodoi{10.1093/mnras/stz3169}

\bibitem[{A.~M. {Cody} \& D.~D. {Sasselov}(2005){Cody} \& {Sasselov}}]{Cody2005}
{Cody}, A.~M., \& {Sasselov}, D.~D. 2005, \bibinfo{title}{{Stellar Evolution with Enriched Surface Convection Zones. I. General Effects of Planet Consumption},} \apj, 622, 704, \dodoi{10.1086/427909}

\bibitem[{R.~H. {Cyburt} {et~al.}(2010){Cyburt}, {Amthor}, {Ferguson}, {Meisel}, {Smith}, {Warren}, {Heger}, {Hoffman}, {Rauscher}, {Sakharuk}, {Schatz}, {Thielemann}, \& {Wiescher}}]{Cyburt2010}
{Cyburt}, R.~H., {Amthor}, A.~M., {Ferguson}, R., {et~al.} 2010, \bibinfo{title}{{The JINA REACLIB Database: Its Recent Updates and Impact on Type-I X-ray Bursts},} \apjs, 189, 240, \dodoi{10.1088/0067-0049/189/1/240}

\bibitem[{C. {Damiani} \& R.~F. {D{\'\i}az}(2016){Damiani} \& {D{\'\i}az}}]{Damiani2016}
{Damiani}, C., \& {D{\'\i}az}, R.~F. 2016, \bibinfo{title}{{Can brown dwarfs survive on close orbits around convective stars?},} \aap, 589, A55, \dodoi{10.1051/0004-6361/201527100}

\bibitem[{K. {De} {et~al.}(2023){De}, {MacLeod}, {Karambelkar}, {Jencson}, {Chakrabarty}, {Conroy}, {Dekany}, {Eilers}, {Graham}, {Hillenbrand}, {Kara}, {Kasliwal}, {Kulkarni}, {Lau}, {Loeb}, {Masci}, {Medford}, {Meisner}, {Patel}, {Quiroga-Nu{\~n}ez}, {Riddle}, {Rusholme}, {Simcoe}, {Sjouwerman}, {Teague}, \& {Vanderburg}}]{De2023}
{De}, K., {MacLeod}, M., {Karambelkar}, V., {et~al.} 2023, \bibinfo{title}{{An infrared transient from a star engulfing a planet},} \nat, 617, 55, \dodoi{10.1038/s41586-023-05842-x}

\bibitem[{J.~A. {Faber} {et~al.}(2005){Faber}, {Rasio}, \& {Willems}}]{Faber2005}
{Faber}, J.~A., {Rasio}, F.~A., \& {Willems}, B. 2005, \bibinfo{title}{{Tidal interactions and disruptions of giant planets on highly eccentric orbits},} \icarus, 175, 248, \dodoi{10.1016/j.icarus.2004.10.021}

\bibitem[{J.~W. {Ferguson} {et~al.}(2005){Ferguson}, {Alexander}, {Allard}, {Barman}, {Bodnarik}, {Hauschildt}, {Heffner-Wong}, \& {Tamanai}}]{Ferguson2005}
{Ferguson}, J.~W., {Alexander}, D.~R., {Allard}, F., {et~al.} 2005, \bibinfo{title}{{Low-Temperature Opacities},} \apj, 623, 585, \dodoi{10.1086/428642}

\bibitem[{G.~M. {Fuller} {et~al.}(1985){Fuller}, {Fowler}, \& {Newman}}]{Fuller1985}
{Fuller}, G.~M., {Fowler}, W.~A., \& {Newman}, M.~J. 1985, \bibinfo{title}{{Stellar weak interaction rates for intermediate-mass nuclei. IV - Interpolation procedures for rapidly varying lepton capture rates using effective log (ft)-values},} \apj, 293, 1, \dodoi{10.1086/163208}

\bibitem[{ {Gaia Collaboration} {et~al.}(2016){Gaia Collaboration}, {Prusti}, {de Bruijne}, {Brown}, {Vallenari}, {Babusiaux}, {Bailer-Jones}, {Bastian}, {Biermann}, {Evans}, {Eyer}, {Jansen}, {Jordi}, {Klioner}, {Lammers}, {Lindegren}, {Luri}, {Mignard}, {Milligan}, {Panem}, {Poinsignon}, {Pourbaix}, {Randich}, {Sarri}, {Sartoretti}, {Siddiqui}, {Soubiran}, {Valette}, {van Leeuwen}, {Walton}, {Aerts}, {Arenou}, {Cropper}, {Drimmel}, {H{\o}g}, {Katz}, {Lattanzi}, {O'Mullane}, {Grebel}, {Holland}, {Huc}, {Passot}, {Bramante}, {Cacciari}, {Casta{\~n}eda}, {Chaoul}, {Cheek}, {De Angeli}, {Fabricius}, {Guerra}, {Hern{\'a}ndez}, {Jean-Antoine-Piccolo}, {Masana}, {Messineo}, {Mowlavi}, {Nienartowicz}, {Ord{\'o}{\~n}ez-Blanco}, {Panuzzo}, {Portell}, {Richards}, {Riello}, {Seabroke}, {Tanga}, {Th{\'e}venin}, {Torra}, {Els}, {Gracia-Abril}, {Comoretto}, {Garcia-Reinaldos}, {Lock}, {Mercier}, {Altmann}, {Andrae}, {Astraatmadja}, {Bellas-Velidis}, {Benson}, {Berthier}, {Blomme}, {Busso}, {Carry}, {Cellino}, {Clementini}, {Cowell}, {Creevey}, {Cuypers}, {Davidson}, {De Ridder}, {de Torres}, {Delchambre}, {Dell'Oro}, {Ducourant}, {Fr{\'e}mat}, {Garc{\'\i}a-Torres}, {Gosset}, {Halbwachs}, {Hambly}, {Harrison}, {Hauser}, {Hestroffer}, {Hodgkin}, {Huckle}, {Hutton}, {Jasniewicz}, {Jordan}, {Kontizas}, {Korn}, {Lanzafame}, {Manteiga}, {Moitinho}, {Muinonen}, {Osinde}, {Pancino}, {Pauwels}, {Petit}, {Recio-Blanco}, {Robin}, {Sarro}, {Siopis}, {Smith}, {Smith}, {Sozzetti}, {Thuillot}, {van Reeven}, {Viala}, {Abbas}, {Abreu Aramburu}, {Accart}, {Aguado}, {Allan}, {Allasia}, {Altavilla}, {{\'A}lvarez}, {Alves}, {Anderson}, {Andrei}, {Anglada Varela}, {Antiche}, {Antoja}, {Ant{\'o}n}, {Arcay}, {Atzei}, {Ayache}, {Bach}, {Baker}, {Balaguer-N{\'u}{\~n}ez}, {Barache}, {Barata}, {Barbier}, {Barblan}, {Baroni}, {Barrado y Navascu{\'e}s}, {Barros}, {Barstow}, {Becciani}, {Bellazzini}, {Bellei}, {Bello Garc{\'\i}a}, {Belokurov}, {Bendjoya}, {Berihuete}, {Bianchi}, {Bienaym{\'e}}, {Billebaud}, {Blagorodnova}, {Blanco-Cuaresma}, {Boch}, {Bombrun}, {Borrachero}, {Bouquillon}, {Bourda}, {Bouy}, {Bragaglia}, {Breddels}, {Brouillet}, {Br{\"u}semeister}, {Bucciarelli}, {Budnik}, {Burgess}, {Burgon}, {Burlacu}, {Busonero}, {Buzzi}, {Caffau}, {Cambras}, {Campbell}, {Cancelliere}, {Cantat-Gaudin}, {Carlucci}, {Carrasco}, {Castellani}, {Charlot}, {Charnas}, {Charvet}, {Chassat}, {Chiavassa}, {Clotet}, {Cocozza}, {Collins}, {Collins}, \& {Costigan}}]{GaiaCollaboration2016}
{Gaia Collaboration}, {Prusti}, T., {de Bruijne}, J.~H.~J., {et~al.} 2016, \bibinfo{title}{{The Gaia mission},} \aap, 595, A1, \dodoi{10.1051/0004-6361/201629272}

\bibitem[{ {Gaia Collaboration} {et~al.}(2021){Gaia Collaboration}, {Brown}, {Vallenari}, {Prusti}, {de Bruijne}, {Babusiaux}, {Biermann}, {Creevey}, {Evans}, {Eyer}, {Hutton}, {Jansen}, {Jordi}, {Klioner}, {Lammers}, {Lindegren}, {Luri}, {Mignard}, {Panem}, {Pourbaix}, {Randich}, {Sartoretti}, {Soubiran}, {Walton}, {Arenou}, {Bailer-Jones}, {Bastian}, {Cropper}, {Drimmel}, {Katz}, {Lattanzi}, {van Leeuwen}, {Bakker}, {Cacciari}, {Casta{\~n}eda}, {De Angeli}, {Ducourant}, {Fabricius}, {Fouesneau}, {Fr{\'e}mat}, {Guerra}, {Guerrier}, {Guiraud}, {Jean-Antoine Piccolo}, {Masana}, {Messineo}, {Mowlavi}, {Nicolas}, {Nienartowicz}, {Pailler}, {Panuzzo}, {Riclet}, {Roux}, {Seabroke}, {Sordo}, {Tanga}, {Th{\'e}venin}, {Gracia-Abril}, {Portell}, {Teyssier}, {Altmann}, {Andrae}, {Bellas-Velidis}, {Benson}, {Berthier}, {Blomme}, {Brugaletta}, {Burgess}, {Busso}, {Carry}, {Cellino}, {Cheek}, {Clementini}, {Damerdji}, {Davidson}, {Delchambre}, {Dell'Oro}, {Fern{\'a}ndez-Hern{\'a}ndez}, {Galluccio}, {Garc{\'\i}a-Lario}, {Garcia-Reinaldos}, {Gonz{\'a}lez-N{\'u}{\~n}ez}, {Gosset}, {Haigron}, {Halbwachs}, {Hambly}, {Harrison}, {Hatzidimitriou}, {Heiter}, {Hern{\'a}ndez}, {Hestroffer}, {Hodgkin}, {Holl}, {Jan{\ss}en}, {Jevardat de Fombelle}, {Jordan}, {Krone-Martins}, {Lanzafame}, {L{\"o}ffler}, {Lorca}, {Manteiga}, {Marchal}, {Marrese}, {Moitinho}, {Mora}, {Muinonen}, {Osborne}, {Pancino}, {Pauwels}, {Petit}, {Recio-Blanco}, {Richards}, {Riello}, {Rimoldini}, {Robin}, {Roegiers}, {Rybizki}, {Sarro}, {Siopis}, {Smith}, {Sozzetti}, {Ulla}, {Utrilla}, {van Leeuwen}, {van Reeven}, {Abbas}, {Abreu Aramburu}, {Accart}, {Aerts}, {Aguado}, {Ajaj}, {Altavilla}, {{\'A}lvarez}, {{\'A}lvarez Cid-Fuentes}, {Alves}, {Anderson}, {Anglada Varela}, {Antoja}, {Audard}, {Baines}, {Baker}, {Balaguer-N{\'u}{\~n}ez}, {Balbinot}, {Balog}, {Barache}, {Barbato}, {Barros}, {Barstow}, {Bartolom{\'e}}, {Bassilana}, {Bauchet}, {Baudesson-Stella}, {Becciani}, {Bellazzini}, {Bernet}, {Bertone}, {Bianchi}, {Blanco-Cuaresma}, {Boch}, {Bombrun}, {Bossini}, {Bouquillon}, {Bragaglia}, {Bramante}, {Breedt}, {Bressan}, {Brouillet}, {Bucciarelli}, {Burlacu}, {Busonero}, {Butkevich}, {Buzzi}, {Caffau}, {Cancelliere}, {C{\'a}novas}, {Cantat-Gaudin}, {Carballo}, {Carlucci}, {Carnerero}, {Carrasco}, {Casamiquela}, {Castellani}, {Castro-Ginard}, {Castro Sampol}, {Chaoul}, {Charlot}, {Chemin}, {Chiavassa}, {Cioni}, {Comoretto}, {Cooper}, {Cornez}, {Cowell}, {Crifo}, {Crosta}, {Crowley}, {Dafonte}, {Dapergolas}, {David}, \& {David}}]{GaiaCollaboration2021}
{Gaia Collaboration}, {Brown}, A.~G.~A., {Vallenari}, A., {et~al.} 2021, \bibinfo{title}{{Gaia Early Data Release 3. Summary of the contents and survey properties},} \aap, 649, A1, \dodoi{10.1051/0004-6361/202039657}

\bibitem[{N.~J. Goldbaum {et~al.}(2018)Goldbaum, ZuHone, Turk, Kowalik, \& Rosen}]{Goldbaum2018}
Goldbaum, N.~J., ZuHone, J.~A., Turk, M.~J., Kowalik, K., \& Rosen, A.~L. 2018, \bibinfo{title}{unyt: Handle, manipulate, and convert data with units in Python,} Journal of Open Source Software, 3, 809, \dodoi{10.21105/joss.00809}

\bibitem[{R.~G. {Gratton} {et~al.}(2001){Gratton}, {Bonanno}, {Claudi}, {Cosentino}, {Desidera}, {Lucatello}, \& {Scuderi}}]{Gratton2001}
{Gratton}, R.~G., {Bonanno}, G., {Claudi}, R.~U., {et~al.} 2001, \bibinfo{title}{{Non-interacting main-sequence binaries with different chemical compositions: Evidences of infall of rocky material?},} \aap, 377, 123, \dodoi{10.1051/0004-6361:20011066}

\bibitem[{J. {Guillochon} {et~al.}(2011){Guillochon}, {Ramirez-Ruiz}, \& {Lin}}]{Guillochon2011}
{Guillochon}, J., {Ramirez-Ruiz}, E., \& {Lin}, D. 2011, \bibinfo{title}{{Consequences of the Ejection and Disruption of Giant Planets},} \apj, 732, 74, \dodoi{10.1088/0004-637X/732/2/74}

\bibitem[{S.-S. {Guo}(2023){Guo}}]{Guo2023}
{Guo}, S.-S. 2023, \bibinfo{title}{{The Impact of Tidal Migration of Hot Jupiters on the Rotation of Sun-like Main-sequence Stars},} Research in Astronomy and Astrophysics, 23, 095014, \dodoi{10.1088/1674-4527/ace028}

\bibitem[{J.~H. {Hamer} \& K.~C. {Schlaufman}(2019){Hamer} \& {Schlaufman}}]{Hamer2019}
{Hamer}, J.~H., \& {Schlaufman}, K.~C. 2019, \bibinfo{title}{{Hot Jupiters Are Destroyed by Tides While Their Host Stars Are on the Main Sequence},} \aj, 158, 190, \dodoi{10.3847/1538-3881/ab3c56}

\bibitem[{C.~R. Harris {et~al.}(2020)Harris, Millman, van~der Walt, Gommers, Virtanen, Cournapeau, Wieser, Taylor, Berg, Smith, Kern, Picus, Hoyer, van Kerkwijk, Brett, Haldane, Fernández~del Río, Wiebe, Peterson, Gérard-Marchant, Sheppard, Reddy, Weckesser, Abbasi, Gohlke, \& Oliphant}]{Harris2020}
Harris, C.~R., Millman, K.~J., van~der Walt, S.~J., {et~al.} 2020, \bibinfo{title}{Array programming with {NumPy},} Nature, 585, 357–362, \dodoi{10.1038/s41586-020-2649-2}

\bibitem[{J.~D. {Hartman} {et~al.}(2016){Hartman}, {Bakos}, {Bhatti}, {Penev}, {Bieryla}, {Latham}, {Kov{\'a}cs}, {Torres}, {Csubry}, {de Val-Borro}, {Buchhave}, {Kov{\'a}cs}, {Quinn}, {Howard}, {Isaacson}, {Fulton}, {Everett}, {Esquerdo}, {B{\'e}ky}, {Szklenar}, {Falco}, {Santerne}, {Boisse}, {H{\'e}brard}, {Burrows}, {L{\'a}z{\'a}r}, {Papp}, \& {S{\'a}ri}}]{Hartman2016}
{Hartman}, J.~D., {Bakos}, G.~{\'A}., {Bhatti}, W., {et~al.} 2016, \bibinfo{title}{{HAT-P-65b and HAT-P-66b: Two Transiting Inflated Hot Jupiters and Observational Evidence for the Reinflation of Close-in Giant Planets},} \aj, 152, 182, \dodoi{10.3847/0004-6256/152/6/182}

\bibitem[{L. {Hartmann} {et~al.}(2016){Hartmann}, {Herczeg}, \& {Calvet}}]{Hartmann2016}
{Hartmann}, L., {Herczeg}, G., \& {Calvet}, N. 2016, \bibinfo{title}{{Accretion onto Pre-Main-Sequence Stars},} \araa, 54, 135, \dodoi{10.1146/annurev-astro-081915-023347}

\bibitem[{G.~J. {Herczeg} {et~al.}(2016){Herczeg}, {Dong}, {Shappee}, {Chen}, {Hillenbrand}, {Jose}, {Kochanek}, {Prieto}, {Stanek}, {Kaplan}, {Holoien}, {Mairs}, {Johnstone}, {Gully-Santiago}, {Zhu}, {Smith}, {Bersier}, {Mulders}, {Filippenko}, {Ayani}, {Brimacombe}, {Brown}, {Connelley}, {Harmanen}, {Itoh}, {Kawabata}, {Maehara}, {Takata}, {Yuk}, \& {Zheng}}]{Herczeg2016}
{Herczeg}, G.~J., {Dong}, S., {Shappee}, B.~J., {et~al.} 2016, \bibinfo{title}{{The Eruption of the Candidate Young Star ASASSN-15QI},} \apj, 831, 133, \dodoi{10.3847/0004-637X/831/2/133}

\bibitem[{A.~W. {Howard} {et~al.}(2012){Howard}, {Marcy}, {Bryson}, {Jenkins}, {Rowe}, {Batalha}, {Borucki}, {Koch}, {Dunham}, {Gautier}, {Van Cleve}, {Cochran}, {Latham}, {Lissauer}, {Torres}, {Brown}, {Gilliland}, {Buchhave}, {Caldwell}, {Christensen-Dalsgaard}, {Ciardi}, {Fressin}, {Haas}, {Howell}, {Kjeldsen}, {Seager}, {Rogers}, {Sasselov}, {Steffen}, {Basri}, {Charbonneau}, {Christiansen}, {Clarke}, {Dupree}, {Fabrycky}, {Fischer}, {Ford}, {Fortney}, {Tarter}, {Girouard}, {Holman}, {Johnson}, {Klaus}, {Machalek}, {Moorhead}, {Morehead}, {Ragozzine}, {Tenenbaum}, {Twicken}, {Quinn}, {Isaacson}, {Shporer}, {Lucas}, {Walkowicz}, {Welsh}, {Boss}, {Devore}, {Gould}, {Smith}, {Morris}, {Prsa}, {Morton}, {Still}, {Thompson}, {Mullally}, {Endl}, \& {MacQueen}}]{Howard2012}
{Howard}, A.~W., {Marcy}, G.~W., {Bryson}, S.~T., {et~al.} 2012, \bibinfo{title}{{Planet Occurrence within 0.25 AU of Solar-type Stars from Kepler},} \apjs, 201, 15, \dodoi{10.1088/0067-0049/201/2/15}

\bibitem[{J.~D. Hunter(2007)Hunter}]{Hunter2007}
Hunter, J.~D. 2007, \bibinfo{title}{Matplotlib: A 2D graphics environment,} Computing in Science \& Engineering, 9, 90, \dodoi{10.1109/MCSE.2007.55}

\bibitem[{T. {Hutchinson-Smith} {et~al.}(2024){Hutchinson-Smith}, {Everson}, {Twum}, {Batta}, {Yarza}, {Law-Smith}, {Vigna-G{\'o}mez}, \& {Ramirez-Ruiz}}]{HutchinsonSmith2024}
{Hutchinson-Smith}, T., {Everson}, R.~W., {Twum}, A.~A., {et~al.} 2024, \bibinfo{title}{{Rethinking Thorne{\textendash}{\.Z}ytkow Object Formation: The Fate of X-Ray Binary LMC X-4 and Implications for Ultra-long Gamma-Ray Bursts},} \apj, 977, 196, \dodoi{10.3847/1538-4357/ad88f3}

\bibitem[{C.~A. {Iglesias} \& F.~J. {Rogers}(1993){Iglesias} \& {Rogers}}]{Iglesias1993}
{Iglesias}, C.~A., \& {Rogers}, F.~J. 1993, \bibinfo{title}{{Radiative opacities for carbon- and oxygen-rich mixtures},} \apj, 412, 752, \dodoi{10.1086/172958}

\bibitem[{C.~A. {Iglesias} \& F.~J. {Rogers}(1996){Iglesias} \& {Rogers}}]{Iglesias1996}
{Iglesias}, C.~A., \& {Rogers}, F.~J. 1996, \bibinfo{title}{{Updated Opal Opacities},} \apj, 464, 943, \dodoi{10.1086/177381}

\bibitem[{A.~W. {Irwin}(2004){Irwin}}]{Irwin2004}
{Irwin}, A.~W. 2004, \bibinfo{title}{The FreeEOS Code for Calculating the Equation of State for Stellar Interiors,} \url{http://freeeos.sourceforge.net/}

\bibitem[{G. {Israelian} {et~al.}(2001){Israelian}, {Santos}, {Mayor}, \& {Rebolo}}]{Israelian2001}
{Israelian}, G., {Santos}, N.~C., {Mayor}, M., \& {Rebolo}, R. 2001, \bibinfo{title}{{Evidence for planet engulfment by the star HD82943},} \nat, 411, 163, \dodoi{10.1038/35075512}

\bibitem[{N. {Itoh} {et~al.}(1996){Itoh}, {Hayashi}, {Nishikawa}, \& {Kohyama}}]{Itoh1996}
{Itoh}, N., {Hayashi}, H., {Nishikawa}, A., \& {Kohyama}, Y. 1996, \bibinfo{title}{{Neutrino Energy Loss in Stellar Interiors. VII. Pair, Photo-, Plasma, Bremsstrahlung, and Recombination Neutrino Processes},} \apjs, 102, 411, \dodoi{10.1086/192264}

\bibitem[{N. {Ivanova} {et~al.}(2013){Ivanova}, {Justham}, {Avendano Nandez}, \& {Lombardi}}]{Ivanova2013}
{Ivanova}, N., {Justham}, S., {Avendano Nandez}, J.~L., \& {Lombardi}, J.~C. 2013, \bibinfo{title}{{Identification of the Long-Sought Common-Envelope Events},} Science, 339, 433, \dodoi{10.1126/science.1225540}

\bibitem[{B. {Jackson} {et~al.}(2009){Jackson}, {Barnes}, \& {Greenberg}}]{Jackson2009}
{Jackson}, B., {Barnes}, R., \& {Greenberg}, R. 2009, \bibinfo{title}{{Observational Evidence for Tidal Destruction of Exoplanets},} \apj, 698, 1357, \dodoi{10.1088/0004-637X/698/2/1357}

\bibitem[{B. {Jackson} {et~al.}(2016){Jackson}, {Jensen}, {Peacock}, {Arras}, \& {Penev}}]{Jackson2016}
{Jackson}, B., {Jensen}, E., {Peacock}, S., {Arras}, P., \& {Penev}, K. 2016, \bibinfo{title}{{Tidal decay and stable Roche-lobe overflow of short-period gaseous exoplanets},} Celestial Mechanics and Dynamical Astronomy, 126, 227, \dodoi{10.1007/s10569-016-9704-1}

\bibitem[{A.~S. {Jermyn} {et~al.}(2021){Jermyn}, {Schwab}, {Bauer}, {Timmes}, \& {Potekhin}}]{Jermyn2021}
{Jermyn}, A.~S., {Schwab}, J., {Bauer}, E., {Timmes}, F.~X., \& {Potekhin}, A.~Y. 2021, \bibinfo{title}{{Skye: A Differentiable Equation of State},} \apj, 913, 72, \dodoi{10.3847/1538-4357/abf48e}

\bibitem[{A.~S. {Jermyn} {et~al.}(2023){Jermyn}, {Bauer}, {Schwab}, {Farmer}, {Ball}, {Bellinger}, {Dotter}, {Joyce}, {Marchant}, {Mombarg}, {Wolf}, {Sunny Wong}, {Cinquegrana}, {Farrell}, {Smolec}, {Thoul}, {Cantiello}, {Herwig}, {Toloza}, {Bildsten}, {Townsend}, \& {Timmes}}]{Jermyn2023}
{Jermyn}, A.~S., {Bauer}, E.~B., {Schwab}, J., {et~al.} 2023, \bibinfo{title}{{Modules for Experiments in Stellar Astrophysics (MESA): Time-dependent Convection, Energy Conservation, Automatic Differentiation, and Infrastructure},} \apjs, 265, 15, \dodoi{10.3847/1538-4365/acae8d}

\bibitem[{S. {Jia} \& H.~C. {Spruit}(2018){Jia} \& {Spruit}}]{Jia2018}
{Jia}, S., \& {Spruit}, H.~C. 2018, \bibinfo{title}{{Disruption of a Planet Spiraling into its Host Star},} \apj, 864, 169, \dodoi{10.3847/1538-4357/aad77c}

\bibitem[{S.~R. {Kane}(2023){Kane}}]{Kane2023}
{Kane}, S.~R. 2023, \bibinfo{title}{{Planetary Engulfment Prognosis within the {\ensuremath{\rho}} CrB System},} \apj, 958, 120, \dodoi{10.3847/1538-4357/ad06b2}

\bibitem[{A. {Kashi}(2018){Kashi}}]{Kashi2018}
{Kashi}, A. 2018, \bibinfo{title}{{Simulations and Modeling of Intermediate Luminosity Optical Transients and Supernova Impostors},} Galaxies, 6, 82, \dodoi{10.3390/galaxies6030082}

\bibitem[{A. {Kashi} {et~al.}(2019){Kashi}, {Michaelis}, \& {Feigin}}]{Kashi2019}
{Kashi}, A., {Michaelis}, A.~M., \& {Feigin}, L. 2019, \bibinfo{title}{{ASASSN-13db 2014{\textendash}2017 Eruption as an Intermediate Luminosity Optical Transient},} Galaxies, 8, 2, \dodoi{10.3390/galaxies8010002}

\bibitem[{A. {Kashi} \& N. {Soker}(2017){Kashi} \& {Soker}}]{Kashi2017}
{Kashi}, A., \& {Soker}, N. 2017, \bibinfo{title}{{An intermediate luminosity optical transient (ILOTs) model for the young stellar object ASASSN-15qi},} \mnras, 468, 4938, \dodoi{10.1093/mnras/stx767}

\bibitem[{T.~D. {Komacek} \& A.~N. {Youdin}(2017){Komacek} \& {Youdin}}]{Komacek2017}
{Komacek}, T.~D., \& {Youdin}, A.~N. 2017, \bibinfo{title}{{Structure and Evolution of Internally Heated Hot Jupiters},} \apj, 844, 94, \dodoi{10.3847/1538-4357/aa7b75}

\bibitem[{M. {Kramer} {et~al.}(2020){Kramer}, {Schneider}, {Ohlmann}, {Geier}, {Schaffenroth}, {Pakmor}, \& {R{\"o}pke}}]{Kramer2020}
{Kramer}, M., {Schneider}, F.~R.~N., {Ohlmann}, S.~T., {et~al.} 2020, \bibinfo{title}{{Formation of sdB-stars via common envelope ejection by substellar companions},} \aap, 642, A97, \dodoi{10.1051/0004-6361/202038702}

\bibitem[{M. {Kunitomo} {et~al.}(2011){Kunitomo}, {Ikoma}, {Sato}, {Katsuta}, \& {Ida}}]{Kunitomo2011}
{Kunitomo}, M., {Ikoma}, M., {Sato}, B., {Katsuta}, Y., \& {Ida}, S. 2011, \bibinfo{title}{{Planet Engulfment by \raisebox{-0.5ex}\textasciitilde1.5-3 M $_{sun}$ Red Giants},} \apj, 737, 66, \dodoi{10.1088/0004-637X/737/2/66}

\bibitem[{K. {Langanke} \& G. {Mart{\'\i}nez-Pinedo}(2000){Langanke} \& {Mart{\'\i}nez-Pinedo}}]{Langanke2000}
{Langanke}, K., \& {Mart{\'\i}nez-Pinedo}, G. 2000, \bibinfo{title}{{Shell-model calculations of stellar weak interaction rates: II. Weak rates for nuclei in the mass range /A=45-65 in supernovae environments},} \nphysa, 673, 481, \dodoi{10.1016/S0375-9474(00)00131-7}

\bibitem[{M.~Y.~M. {Lau} {et~al.}(2025){Lau}, {Cantiello}, {Jermyn}, {MacLeod}, {Mandel}, \& {Price}}]{Lau2025}
{Lau}, M. Y.~M., {Cantiello}, M., {Jermyn}, A.~S., {et~al.} 2025, \bibinfo{title}{{Hot Jupiter engulfment by an early red giant in 3D hydrodynamics},} \aap, 694, A264, \dodoi{10.1051/0004-6361/202452081}

\bibitem[{R.~M. {Lau} {et~al.}(2025){Lau}, {Jencson}, {Salyk}, {De}, {Fox}, {Hankins}, {Kasliwal}, {Keyes}, {Macleod}, {Ressler}, \& {Rose}}]{Lau2025a}
{Lau}, R.~M., {Jencson}, J.~E., {Salyk}, C., {et~al.} 2025, \bibinfo{title}{{Revealing a Main-sequence Star that Consumed a Planet with JWST},} \apj, 983, 87, \dodoi{10.3847/1538-4357/adb429}

\bibitem[{G. {Laughlin} \& F.~C. {Adams}(1997){Laughlin} \& {Adams}}]{Laughlin1997}
{Laughlin}, G., \& {Adams}, F.~C. 1997, \bibinfo{title}{{Possible Stellar Metallicity Enhancements from the Accretion of Planets},} \apjl, 491, L51, \dodoi{10.1086/311056}

\bibitem[{A. {Lawrence} {et~al.}(2007){Lawrence}, {Warren}, {Almaini}, {Edge}, {Hambly}, {Jameson}, {Lucas}, {Casali}, {Adamson}, {Dye}, {Emerson}, {Foucaud}, {Hewett}, {Hirst}, {Hodgkin}, {Irwin}, {Lodieu}, {McMahon}, {Simpson}, {Smail}, {Mortlock}, \& {Folger}}]{Lawrence2007}
{Lawrence}, A., {Warren}, S.~J., {Almaini}, O., {et~al.} 2007, \bibinfo{title}{{The UKIRT Infrared Deep Sky Survey (UKIDSS)},} \mnras, 379, 1599, \dodoi{10.1111/j.1365-2966.2007.12040.x}

\bibitem[{Y.~A. {Lazovik}(2023){Lazovik}}]{Lazovik2023}
{Lazovik}, Y.~A. 2023, \bibinfo{title}{{Unravelling the evolution of hot Jupiter systems under the effect of tidal and magnetic interactions and mass-loss},} \mnras, 520, 3749, \dodoi{10.1093/mnras/stad394}

\bibitem[{P. {Leonardi} {et~al.}(2024){Leonardi}, {Nascimbeni}, {Granata}, {Malavolta}, {Borsato}, {Biazzo}, {Lanza}, {Desidera}, {Piotto}, {Nardiello}, {Damasso}, {Cunial}, \& {Bedin}}]{Leonardi2024}
{Leonardi}, P., {Nascimbeni}, V., {Granata}, V., {et~al.} 2024, \bibinfo{title}{{TASTE. V. A new ground-based investigation of orbital decay in the ultra-hot Jupiter WASP-12b},} \aap, 686, A84, \dodoi{10.1051/0004-6361/202348363}

\bibitem[{B. {Levrard} {et~al.}(2009){Levrard}, {Winisdoerffer}, \& {Chabrier}}]{Levrard2009}
{Levrard}, B., {Winisdoerffer}, C., \& {Chabrier}, G. 2009, \bibinfo{title}{{Falling Transiting Extrasolar Giant Planets},} \apjl, 692, L9, \dodoi{10.1088/0004-637X/692/1/L9}

\bibitem[{S.~L. {Li} {et~al.}(2008){Li}, {Lin}, \& {Liu}}]{Li2008}
{Li}, S.~L., {Lin}, D.~N.~C., \& {Liu}, X.~W. 2008, \bibinfo{title}{{Extent of Pollution in Planet-bearing Stars},} \apj, 685, 1210, \dodoi{10.1086/591122}

\bibitem[{F. {Liu} {et~al.}(2024){Liu}, {Ting}, {Yong}, {Bitsch}, {Karakas}, {Murphy}, {Joyce}, {Dotter}, \& {Dai}}]{Liu2024}
{Liu}, F., {Ting}, Y.-S., {Yong}, D., {et~al.} 2024, \bibinfo{title}{{At least one in a dozen stars shows evidence of planetary ingestion},} \nat, 627, 501, \dodoi{10.1038/s41586-024-07091-y}

\bibitem[{S.-F. {Liu} {et~al.}(2013){Liu}, {Guillochon}, {Lin}, \& {Ramirez-Ruiz}}]{Liu2013}
{Liu}, S.-F., {Guillochon}, J., {Lin}, D. N.~C., \& {Ramirez-Ruiz}, E. 2013, \bibinfo{title}{{On the Survivability and Metamorphism of Tidally Disrupted Giant Planets: The Role of Dense Cores},} \apj, 762, 37, \dodoi{10.1088/0004-637X/762/1/37}

\bibitem[{M. {Livio} \& N. {Soker}(2002){Livio} \& {Soker}}]{Livio2002}
{Livio}, M., \& {Soker}, N. 2002, \bibinfo{title}{{The Effects of Planets and Brown Dwarfs on Stellar Rotation and Mass Loss},} \apjl, 571, L161, \dodoi{10.1086/341411}

\bibitem[{E.~D. {Lopez} \& J.~J. {Fortney}(2016){Lopez} \& {Fortney}}]{Lopez2016}
{Lopez}, E.~D., \& {Fortney}, J.~J. 2016, \bibinfo{title}{{Re-inflated Warm Jupiters around Red Giants},} \apj, 818, 4, \dodoi{10.3847/0004-637X/818/1/4}

\bibitem[{G. {Maciejewski} {et~al.}(2016){Maciejewski}, {Dimitrov}, {Fern{\'a}ndez}, {Sota}, {Nowak}, {Ohlert}, {Nikolov}, {Bukowiecki}, {Hinse}, {Pall{\'e}}, {Tingley}, {Kjurkchieva}, {Lee}, \& {Lee}}]{Maciejewski2016}
{Maciejewski}, G., {Dimitrov}, D., {Fern{\'a}ndez}, M., {et~al.} 2016, \bibinfo{title}{{Departure from the constant-period ephemeris for the transiting exoplanet WASP-12},} \aap, 588, L6, \dodoi{10.1051/0004-6361/201628312}

\bibitem[{G. {Maciejewski} {et~al.}(2018){Maciejewski}, {Fern{\'a}ndez}, {Aceituno}, {Mart{\'\i}n-Ruiz}, {Ohlert}, {Dimitrov}, {Szyszka}, {von Essen}, {Mugrauer}, {Bischoff}, {Michel}, {Mallonn}, {Stangret}, \& {Mo{\'z}dzierski}}]{Maciejewski2018}
{Maciejewski}, G., {Fern{\'a}ndez}, M., {Aceituno}, F., {et~al.} 2018, \bibinfo{title}{{Planet-Star Interactions with Precise Transit Timing. I. The Refined Orbital Decay Rate for WASP-12 b and Initial Constraints for HAT-P-23 b, KELT-1 b, KELT-16 b, WASP-33 b and WASP-103 b},} \actaa, 68, 371, \dodoi{10.32023/0001-5237/68.4.4}

\bibitem[{M. {MacLeod} {et~al.}(2018{\natexlab{a}}){MacLeod}, {Cantiello}, \& {Soares-Furtado}}]{MacLeod2018}
{MacLeod}, M., {Cantiello}, M., \& {Soares-Furtado}, M. 2018{\natexlab{a}}, \bibinfo{title}{{Planetary Engulfment in the Hertzsprung-Russell Diagram},} \apjl, 853, L1, \dodoi{10.3847/2041-8213/aaa5fa}

\bibitem[{M. {MacLeod} \& A. {Loeb}(2020){MacLeod} \& {Loeb}}]{MacLeod2020}
{MacLeod}, M., \& {Loeb}, A. 2020, \bibinfo{title}{{Pre-common-envelope Mass Loss from Coalescing Binary Systems},} \apj, 895, 29, \dodoi{10.3847/1538-4357/ab89b6}

\bibitem[{M. {MacLeod} {et~al.}(2017){MacLeod}, {Macias}, {Ramirez-Ruiz}, {Grindlay}, {Batta}, \& {Montes}}]{MacLeod2017a}
{MacLeod}, M., {Macias}, P., {Ramirez-Ruiz}, E., {et~al.} 2017, \bibinfo{title}{{Lessons from the Onset of a Common Envelope Episode: the Remarkable M31 2015 Luminous Red Nova Outburst},} \apj, 835, 282, \dodoi{10.3847/1538-4357/835/2/282}

\bibitem[{M. {MacLeod} {et~al.}(2018{\natexlab{b}}){MacLeod}, {Ostriker}, \& {Stone}}]{MacLeod2018a}
{MacLeod}, M., {Ostriker}, E.~C., \& {Stone}, J.~M. 2018{\natexlab{b}}, \bibinfo{title}{{Runaway Coalescence at the Onset of Common Envelope Episodes},} \apj, 863, 5, \dodoi{10.3847/1538-4357/aacf08}

\bibitem[{A. {Massarotti} {et~al.}(2008){Massarotti}, {Latham}, {Stefanik}, \& {Fogel}}]{Massarotti2008}
{Massarotti}, A., {Latham}, D.~W., {Stefanik}, R.~P., \& {Fogel}, J. 2008, \bibinfo{title}{{Rotational and Radial Velocities for a Sample of 761 HIPPARCOS Giants and the Role of Binarity},} \aj, 135, 209, \dodoi{10.1088/0004-6256/135/1/209}

\bibitem[{T. {Matsakos} \& A. {K{\"o}nigl}(2015){Matsakos} \& {K{\"o}nigl}}]{Matsakos2015}
{Matsakos}, T., \& {K{\"o}nigl}, A. 2015, \bibinfo{title}{{A Hot Jupiter for Breakfast? Early Stellar Ingestion of Planets May Be Common},} \apjl, 809, L20, \dodoi{10.1088/2041-8205/809/2/L20}

\bibitem[{T. {Matsumoto} \& B.~D. {Metzger}(2022){Matsumoto} \& {Metzger}}]{Matsumoto2022}
{Matsumoto}, T., \& {Metzger}, B.~D. 2022, \bibinfo{title}{{Light-curve Model for Luminous Red Novae and Inferences about the Ejecta of Stellar Mergers},} \apj, 938, 5, \dodoi{10.3847/1538-4357/ac6269}

\bibitem[{S. {Matsumura} {et~al.}(2010){Matsumura}, {Peale}, \& {Rasio}}]{Matsumura2010}
{Matsumura}, S., {Peale}, S.~J., \& {Rasio}, F.~A. 2010, \bibinfo{title}{{Tidal Evolution of Close-in Planets},} \apj, 725, 1995, \dodoi{10.1088/0004-637X/725/2/1995}

\bibitem[{W. {McKinney}(2010){McKinney}}]{McKinney2010}
{McKinney}, W. 2010, in {P}roceedings of the 9th {P}ython in {S}cience {C}onference, ed. {S}t\'efan van~der {W}alt \& {J}arrod {M}illman, 56 -- 61, \dodoi{10.25080/Majora-92bf1922-00a}

\bibitem[{B.~D. {Metzger} {et~al.}(2012){Metzger}, {Giannios}, \& {Spiegel}}]{Metzger2012}
{Metzger}, B.~D., {Giannios}, D., \& {Spiegel}, D.~S. 2012, \bibinfo{title}{{Optical and X-ray transients from planet-star mergers},} \mnras, 425, 2778, \dodoi{10.1111/j.1365-2966.2012.21444.x}

\bibitem[{B.~D. {Metzger} {et~al.}(2017){Metzger}, {Shen}, \& {Stone}}]{Metzger2017}
{Metzger}, B.~D., {Shen}, K.~J., \& {Stone}, N. 2017, \bibinfo{title}{{Secular dimming of KIC 8462852 following its consumption of a planet},} \mnras, 468, 4399, \dodoi{10.1093/mnras/stx823}

\bibitem[{S.~C. {Millholland} {et~al.}(2025){Millholland}, {MacLeod}, \& {Xiao}}]{Millholland2025}
{Millholland}, S.~C., {MacLeod}, M., \& {Xiao}, F. 2025, \bibinfo{title}{{Empirical Constraints on Tidal Dissipation in Exoplanet Host Stars},} \apj, 981, 77, \dodoi{10.3847/1538-4357/ada76d}

\bibitem[{P. {Miquelarena} {et~al.}(2024){Miquelarena}, {Saffe}, {Flores}, {Petrucci}, {Yana Galarza}, {Alacoria}, {Jaque Arancibia}, {Jofr{\'e}}, {Montenegro Armijo}, \& {Gunella}}]{Miquelarena2024}
{Miquelarena}, P., {Saffe}, C., {Flores}, M., {et~al.} 2024, \bibinfo{title}{{The largest metallicity difference in twin systems: High-precision abundance analysis of the benchmark pair Krios and Kronos},} \aap, 688, A73, \dodoi{10.1051/0004-6361/202449983}

\bibitem[{S. {Miyazaki} \& K. {Masuda}(2023){Miyazaki} \& {Masuda}}]{Miyazaki2023}
{Miyazaki}, S., \& {Masuda}, K. 2023, \bibinfo{title}{{Evidence That the Occurrence Rate of Hot Jupiters around Sun-like Stars Decreases with Stellar Age},} \aj, 166, 209, \dodoi{10.3847/1538-3881/acff71}

\bibitem[{J. {Montalb{\'a}n} \& R. {Rebolo}(2002){Montalb{\'a}n} \& {Rebolo}}]{Montalban2002}
{Montalb{\'a}n}, J., \& {Rebolo}, R. 2002, \bibinfo{title}{{Planet accretion and the abundances of lithium isotopes},} \aap, 386, 1039, \dodoi{10.1051/0004-6361:20020338}

\bibitem[{S.~D. {Murray} {et~al.}(1993){Murray}, {White}, {Blondin}, \& {Lin}}]{Murray1993}
{Murray}, S.~D., {White}, S. D.~M., {Blondin}, J.~M., \& {Lin}, D. N.~C. 1993, \bibinfo{title}{{Dynamical Instabilities in Two-Phase Media and the Minimum Masses of Stellar Systems},} \apj, 407, 588, \dodoi{10.1086/172540}

\bibitem[{R.~A. {Murray-Clay} {et~al.}(2009){Murray-Clay}, {Chiang}, \& {Murray}}]{MurrayClay2009}
{Murray-Clay}, R.~A., {Chiang}, E.~I., \& {Murray}, N. 2009, \bibinfo{title}{{Atmospheric Escape From Hot Jupiters},} \apj, 693, 23, \dodoi{10.1088/0004-637X/693/1/23}

\bibitem[{A. {Mustill}(2024){Mustill}}]{Mustill2024}
{Mustill}, A. 2024, \bibinfo{title}{{Giant branch planetary systems: Dynamical and radiative evolution},} arXiv e-prints, arXiv:2405.09399, \dodoi{10.48550/arXiv.2405.09399}

\bibitem[{A.~J. {Mustill} \& E. {Villaver}(2012){Mustill} \& {Villaver}}]{Mustill2012}
{Mustill}, A.~J., \& {Villaver}, E. 2012, \bibinfo{title}{{Foretellings of Ragnar{\"o}k: World-engulfing Asymptotic Giants and the Inheritance of White Dwarfs},} \apj, 761, 121, \dodoi{10.1088/0004-637X/761/2/121}

\bibitem[{T. {Nagar} {et~al.}(2020){Nagar}, {Spina}, \& {Karakas}}]{Nagar2020}
{Nagar}, T., {Spina}, L., \& {Karakas}, A.~I. 2020, \bibinfo{title}{{The Chemical Signatures of Planetary Engulfment Events in Binary Systems},} \apjl, 888, L9, \dodoi{10.3847/2041-8213/ab5dc6}

\bibitem[{J. {Nordhaus} \& D.~S. {Spiegel}(2013){Nordhaus} \& {Spiegel}}]{Nordhaus2013}
{Nordhaus}, J., \& {Spiegel}, D.~S. 2013, \bibinfo{title}{{On the orbits of low-mass companions to white dwarfs and the fates of the known exoplanets},} \mnras, 432, 500, \dodoi{10.1093/mnras/stt569}

\bibitem[{J. {Nordhaus} {et~al.}(2010){Nordhaus}, {Spiegel}, {Ibgui}, {Goodman}, \& {Burrows}}]{Nordhaus2010}
{Nordhaus}, J., {Spiegel}, D.~S., {Ibgui}, L., {Goodman}, J., \& {Burrows}, A. 2010, \bibinfo{title}{{Tides and tidal engulfment in post-main-sequence binaries: period gaps for planets and brown dwarfs around white dwarfs},} \mnras, 408, 631, \dodoi{10.1111/j.1365-2966.2010.17155.x}

\bibitem[{C.~E. {O'Connor} {et~al.}(2023){O'Connor}, {Bildsten}, {Cantiello}, \& {Lai}}]{OConnor2023}
{O'Connor}, C.~E., {Bildsten}, L., {Cantiello}, M., \& {Lai}, D. 2023, \bibinfo{title}{{Giant Planet Engulfment by Evolved Giant Stars: Light Curves, Asteroseismology, and Survivability},} \apj, 950, 128, \dodoi{10.3847/1538-4357/acd2d4}

\bibitem[{C.~E. {O'Connor} \& D. {Lai}(2025){O'Connor} \& {Lai}}]{OConnor2025}
{O'Connor}, C.~E., \& {Lai}, D. 2025, \bibinfo{title}{{Metal Pollution in Sun-like Stars from Destruction of Ultra{\textendash}short-period Planets},} \apjl, 978, L26, \dodoi{10.3847/2041-8213/ada1ce}

\bibitem[{T. {Oda} {et~al.}(1994){Oda}, {Hino}, {Muto}, {Takahara}, \& {Sato}}]{Oda1994}
{Oda}, T., {Hino}, M., {Muto}, K., {Takahara}, M., \& {Sato}, K. 1994, \bibinfo{title}{{Rate Tables for the Weak Processes of sd-Shell Nuclei in Stellar Matter},} Atomic Data and Nuclear Data Tables, 56, 231, \dodoi{10.1006/adnd.1994.1007}

\bibitem[{A. {Oetjens} {et~al.}(2020){Oetjens}, {Carone}, {Bergemann}, \& {Serenelli}}]{Oetjens2020}
{Oetjens}, A., {Carone}, L., {Bergemann}, M., \& {Serenelli}, A. 2020, \bibinfo{title}{{The influence of planetary engulfment on stellar rotation in metal-poor main-sequence stars},} \aap, 643, A34, \dodoi{10.1051/0004-6361/202038653}

\bibitem[{G.~I. {Ogilvie}(2014){Ogilvie}}]{Ogilvie2014}
{Ogilvie}, G.~I. 2014, \bibinfo{title}{{Tidal Dissipation in Stars and Giant Planets},} \araa, 52, 171, \dodoi{10.1146/annurev-astro-081913-035941}

\bibitem[{S. {Oh} {et~al.}(2018){Oh}, {Price-Whelan}, {Brewer}, {Hogg}, {Spergel}, \& {Myles}}]{Oh2018}
{Oh}, S., {Price-Whelan}, A.~M., {Brewer}, J.~M., {et~al.} 2018, \bibinfo{title}{{Kronos and Krios: Evidence for Accretion of a Massive, Rocky Planetary System in a Comoving Pair of Solar-type Stars},} \apj, 854, 138, \dodoi{10.3847/1538-4357/aaab4d}

\bibitem[{J.-C. {Passy} {et~al.}(2012){Passy}, {Mac Low}, \& {De Marco}}]{Passy2012}
{Passy}, J.-C., {Mac Low}, M.-M., \& {De Marco}, O. 2012, \bibinfo{title}{{On the Survival of Brown Dwarfs and Planets Engulfed by Their Giant Host Star},} \apjl, 759, L30, \dodoi{10.1088/2041-8205/759/2/L30}

\bibitem[{K.~C. {Patra} {et~al.}(2017){Patra}, {Winn}, {Holman}, {Yu}, {Deming}, \& {Dai}}]{Patra2017}
{Patra}, K.~C., {Winn}, J.~N., {Holman}, M.~J., {et~al.} 2017, \bibinfo{title}{{The Apparently Decaying Orbit of WASP-12b},} \aj, 154, 4, \dodoi{10.3847/1538-3881/aa6d75}

\bibitem[{B. {Paxton} {et~al.}(2011){Paxton}, {Bildsten}, {Dotter}, {Herwig}, {Lesaffre}, \& {Timmes}}]{Paxton2011}
{Paxton}, B., {Bildsten}, L., {Dotter}, A., {et~al.} 2011, \bibinfo{title}{{Modules for Experiments in Stellar Astrophysics (MESA)},} \apjs, 192, 3, \dodoi{10.1088/0067-0049/192/1/3}

\bibitem[{B. {Paxton} {et~al.}(2013){Paxton}, {Cantiello}, {Arras}, {Bildsten}, {Brown}, {Dotter}, {Mankovich}, {Montgomery}, {Stello}, {Timmes}, \& {Townsend}}]{Paxton2013}
{Paxton}, B., {Cantiello}, M., {Arras}, P., {et~al.} 2013, \bibinfo{title}{{Modules for Experiments in Stellar Astrophysics (MESA): Planets, Oscillations, Rotation, and Massive Stars},} \apjs, 208, 4, \dodoi{10.1088/0067-0049/208/1/4}

\bibitem[{B. {Paxton} {et~al.}(2015){Paxton}, {Marchant}, {Schwab}, {Bauer}, {Bildsten}, {Cantiello}, {Dessart}, {Farmer}, {Hu}, {Langer}, {Townsend}, {Townsley}, \& {Timmes}}]{Paxton2015}
{Paxton}, B., {Marchant}, P., {Schwab}, J., {et~al.} 2015, \bibinfo{title}{{Modules for Experiments in Stellar Astrophysics (MESA): Binaries, Pulsations, and Explosions},} \apjs, 220, 15, \dodoi{10.1088/0067-0049/220/1/15}

\bibitem[{B. {Paxton} {et~al.}(2018){Paxton}, {Schwab}, {Bauer}, {Bildsten}, {Blinnikov}, {Duffell}, {Farmer}, {Goldberg}, {Marchant}, {Sorokina}, {Thoul}, {Townsend}, \& {Timmes}}]{Paxton2018}
{Paxton}, B., {Schwab}, J., {Bauer}, E.~B., {et~al.} 2018, \bibinfo{title}{{Modules for Experiments in Stellar Astrophysics (MESA): Convective Boundaries, Element Diffusion, and Massive Star Explosions},} \apjs, 234, 34, \dodoi{10.3847/1538-4365/aaa5a8}

\bibitem[{B. {Paxton} {et~al.}(2019){Paxton}, {Smolec}, {Schwab}, {Gautschy}, {Bildsten}, {Cantiello}, {Dotter}, {Farmer}, {Goldberg}, {Jermyn}, {Kanbur}, {Marchant}, {Thoul}, {Townsend}, {Wolf}, {Zhang}, \& {Timmes}}]{Paxton2019}
{Paxton}, B., {Smolec}, R., {Schwab}, J., {et~al.} 2019, \bibinfo{title}{{Modules for Experiments in Stellar Astrophysics (MESA): Pulsating Variable Stars, Rotation, Convective Boundaries, and Energy Conservation},} \apjs, 243, 10, \dodoi{10.3847/1538-4365/ab2241}

\bibitem[{K. {Penev} {et~al.}(2018){Penev}, {Bouma}, {Winn}, \& {Hartman}}]{Penev2018}
{Penev}, K., {Bouma}, L.~G., {Winn}, J.~N., \& {Hartman}, J.~D. 2018, \bibinfo{title}{{Empirical Tidal Dissipation in Exoplanet Hosts From Tidal Spin-up},} \aj, 155, 165, \dodoi{10.3847/1538-3881/aaaf71}

\bibitem[{P. {Podsiadlowski}(1996){Podsiadlowski}}]{Podsiadlowski1996}
{Podsiadlowski}, P. 1996, \bibinfo{title}{{The response of tidally heated stars},} \mnras, 279, 1104, \dodoi{10.1093/mnras/279.4.1104}

\bibitem[{A.~V. {Popkov} \& S.~B. {Popov}(2019){Popkov} \& {Popov}}]{Popkov2019}
{Popkov}, A.~V., \& {Popov}, S.~B. 2019, \bibinfo{title}{{The rate of planet-star coalescences due to tides and stellar evolution},} \mnras, 490, 2390, \dodoi{10.1093/mnras/stz2783}

\bibitem[{A.~Y. {Potekhin} \& G. {Chabrier}(2010){Potekhin} \& {Chabrier}}]{Potekhin2010}
{Potekhin}, A.~Y., \& {Chabrier}, G. 2010, \bibinfo{title}{{Thermodynamic Functions of Dense Plasmas: Analytic Approximations for Astrophysical Applications},} Contributions to Plasma Physics, 50, 82, \dodoi{10.1002/ctpp.201010017}

\bibitem[{J. {Poutanen}(2017){Poutanen}}]{Poutanen2017}
{Poutanen}, J. 2017, \bibinfo{title}{{Rosseland and Flux Mean Opacities for Compton Scattering},} \apj, 835, 119, \dodoi{10.3847/1538-4357/835/2/119}

\bibitem[{G. {Privitera} {et~al.}(2016{\natexlab{a}}){Privitera}, {Meynet}, {Eggenberger}, {Vidotto}, {Villaver}, \& {Bianda}}]{Privitera2016a}
{Privitera}, G., {Meynet}, G., {Eggenberger}, P., {et~al.} 2016{\natexlab{a}}, \bibinfo{title}{{Star-planet interactions. II. Is planet engulfment the origin of fast rotating red giants?},} \aap, 593, A128, \dodoi{10.1051/0004-6361/201628758}

\bibitem[{G. {Privitera} {et~al.}(2016{\natexlab{b}}){Privitera}, {Meynet}, {Eggenberger}, {Vidotto}, {Villaver}, \& {Bianda}}]{Privitera2016}
{Privitera}, G., {Meynet}, G., {Eggenberger}, P., {et~al.} 2016{\natexlab{b}}, \bibinfo{title}{{Star-planet interactions. I. Stellar rotation and planetary orbits},} \aap, 591, A45, \dodoi{10.1051/0004-6361/201528044}

\bibitem[{G. {Privitera} {et~al.}(2016{\natexlab{c}}){Privitera}, {Meynet}, {Eggenberger}, {Georgy}, {Ekstr{\"o}m}, {Vidotto}, {Bianda}, {Villaver}, \& {ud-Doula}}]{Privitera2016b}
{Privitera}, G., {Meynet}, G., {Eggenberger}, P., {et~al.} 2016{\natexlab{c}}, \bibinfo{title}{{High surface magnetic field in red giants as a new signature of planet engulfment?},} \aap, 593, L15, \dodoi{10.1051/0004-6361/201629142}

\bibitem[{A. {Qureshi} {et~al.}(2018){Qureshi}, {Naoz}, \& {Shkolnik}}]{Qureshi2018}
{Qureshi}, A., {Naoz}, S., \& {Shkolnik}, E.~L. 2018, \bibinfo{title}{{Signature of Planetary Mergers on Stellar Spins},} \apj, 864, 65, \dodoi{10.3847/1538-4357/aad562}

\bibitem[{I. {Rapoport} {et~al.}(2021){Rapoport}, {Bear}, \& {Soker}}]{Rapoport2021}
{Rapoport}, I., {Bear}, E., \& {Soker}, N. 2021, \bibinfo{title}{{The future influence of six exoplanets on the envelope properties of their parent stars on the giant branches},} \mnras, 506, 468, \dodoi{10.1093/mnras/stab1774}

\bibitem[{F.~A. {Rasio} {et~al.}(1996){Rasio}, {Tout}, {Lubow}, \& {Livio}}]{Rasio1996}
{Rasio}, F.~A., {Tout}, C.~A., {Lubow}, S.~H., \& {Livio}, M. 1996, \bibinfo{title}{{Tidal Decay of Close Planetary Orbits},} \apj, 470, 1187, \dodoi{10.1086/177941}

\bibitem[{I.~N. {Reid} {et~al.}(1991){Reid}, {Brewer}, {Brucato}, {McKinley}, {Maury}, {Mendenhall}, {Mould}, {Mueller}, {Neugebauer}, {Phinney}, {Sargent}, {Schombert}, \& {Thicksten}}]{Reid1991}
{Reid}, I.~N., {Brewer}, C., {Brucato}, R.~J., {et~al.} 1991, \bibinfo{title}{{The Second Palomar Sky Survey},} \pasp, 103, 661, \dodoi{10.1086/132866}

\bibitem[{F.~J. {Rogers} \& A. {Nayfonov}(2002){Rogers} \& {Nayfonov}}]{Rogers2002}
{Rogers}, F.~J., \& {Nayfonov}, A. 2002, \bibinfo{title}{{Updated and Expanded OPAL Equation-of-State Tables: Implications for Helioseismology},} \apj, 576, 1064, \dodoi{10.1086/341894}

\bibitem[{C. {Saffe} {et~al.}(2017){Saffe}, {Jofr{\'e}}, {Martioli}, {Flores}, {Petrucci}, \& {Jaque Arancibia}}]{Saffe2017}
{Saffe}, C., {Jofr{\'e}}, E., {Martioli}, E., {et~al.} 2017, \bibinfo{title}{{Signatures of rocky planet engulfment in HAT-P-4. Implications for chemical tagging studies},} \aap, 604, L4, \dodoi{10.1051/0004-6361/201731430}

\bibitem[{C. {Saffe} {et~al.}(2024){Saffe}, {Miquelarena}, {Alacoria}, {Martioli}, {Flores}, {Jaque Arancibia}, {Angeloni}, {Jofr{\'e}}, {Yana Galarza}, {Gonz{\'a}lez}, \& {Collado}}]{Saffe2024}
{Saffe}, C., {Miquelarena}, P., {Alacoria}, J., {et~al.} 2024, \bibinfo{title}{{Disentangling the origin of chemical differences using GHOST},} \aap, 682, L23, \dodoi{10.1051/0004-6361/202449263}

\bibitem[{E. {Sandquist} {et~al.}(1998){Sandquist}, {Taam}, {Lin}, \& {Burkert}}]{Sandquist1998}
{Sandquist}, E., {Taam}, R.~E., {Lin}, D.~N.~C., \& {Burkert}, A. 1998, \bibinfo{title}{{Planet Consumption and Stellar Metallicity Enhancements},} \apjl, 506, L65, \dodoi{10.1086/311633}

\bibitem[{E.~L. {Sandquist} {et~al.}(2002){Sandquist}, {Dokter}, {Lin}, \& {Mardling}}]{Sandquist2002}
{Sandquist}, E.~L., {Dokter}, J.~J., {Lin}, D.~N.~C., \& {Mardling}, R.~A. 2002, \bibinfo{title}{{A Critical Examination of Li Pollution and Giant-Planet Consumption by a Host Star},} \apj, 572, 1012, \dodoi{10.1086/340452}

\bibitem[{D. {Saumon} {et~al.}(1995){Saumon}, {Chabrier}, \& {van Horn}}]{Saumon1995}
{Saumon}, D., {Chabrier}, G., \& {van Horn}, H.~M. 1995, \bibinfo{title}{{An Equation of State for Low-Mass Stars and Giant Planets},} \apjs, 99, 713, \dodoi{10.1086/192204}

\bibitem[{M. {Sayeed} {et~al.}(2024){Sayeed}, {Ness}, {Montet}, {Cantiello}, {Casey}, {Buder}, {Bedell}, {Breivik}, {Metzger}, {Martell}, \& {McGee-Gold}}]{Sayeed2024}
{Sayeed}, M., {Ness}, M.~K., {Montet}, B.~T., {et~al.} 2024, \bibinfo{title}{{Many Roads Lead to Lithium: Formation Pathways For Lithium-rich Red Giants},} \apj, 964, 42, \dodoi{10.3847/1538-4357/ad1936}

\bibitem[{K.~C. {Schlaufman} \& J.~N. {Winn}(2013){Schlaufman} \& {Winn}}]{Schlaufman2013}
{Schlaufman}, K.~C., \& {Winn}, J.~N. 2013, \bibinfo{title}{{Evidence for the Tidal Destruction of Hot Jupiters by Subgiant Stars},} \apj, 772, 143, \dodoi{10.1088/0004-637X/772/2/143}

\bibitem[{J. Schwab {et~al.}(2024)Schwab, Wolf, Zingale, Jermyn, Guichandut, \& Mocz}]{Schwab2024}
Schwab, J., Wolf, B., Zingale, M., {et~al.} 2024, \bibinfo{title}{wmwolf/py\_mesa\_reader: 0.3.5,}, 0.3.5 Zenodo, \dodoi{10.5281/zenodo.13697200}

\bibitem[{J. {Sevilla} {et~al.}(2022){Sevilla}, {Behmard}, \& {Fuller}}]{Sevilla2022}
{Sevilla}, J., {Behmard}, A., \& {Fuller}, J. 2022, \bibinfo{title}{{Long-term lithium abundance signatures following planetary engulfment},} \mnras, 516, 3354, \dodoi{10.1093/mnras/stac2436}

\bibitem[{A. {Sicilia-Aguilar} {et~al.}(2017){Sicilia-Aguilar}, {Oprandi}, {Froebrich}, {Fang}, {Prieto}, {Stanek}, {Scholz}, {Kochanek}, {Henning}, {Gredel}, {Holoien}, {Rabus}, {Shappee}, {Billington}, {Campbell-White}, \& {Zegmott}}]{SiciliaAguilar2017}
{Sicilia-Aguilar}, A., {Oprandi}, A., {Froebrich}, D., {et~al.} 2017, \bibinfo{title}{{The 2014-2017 outburst of the young star ASASSN-13db. A time-resolved picture of a very-low-mass star between EXors and FUors},} \aap, 607, A127, \dodoi{10.1051/0004-6361/201731263}

\bibitem[{L. {Siess} \& M. {Livio}(1999{\natexlab{a}}){Siess} \& {Livio}}]{Siess1999}
{Siess}, L., \& {Livio}, M. 1999{\natexlab{a}}, \bibinfo{title}{{The accretion of brown dwarfs and planets by giant stars - I. Asymptotic giant branch stars},} \mnras, 304, 925, \dodoi{10.1046/j.1365-8711.1999.02376.x}

\bibitem[{L. {Siess} \& M. {Livio}(1999{\natexlab{b}}){Siess} \& {Livio}}]{Siess1999a}
{Siess}, L., \& {Livio}, M. 1999{\natexlab{b}}, \bibinfo{title}{{The accretion of brown dwarfs and planets by giant stars - II. Solar-mass stars on the red giant branch},} \mnras, 308, 1133, \dodoi{10.1046/j.1365-8711.1999.02784.x}

\bibitem[{B.~M.~T.~B. {Soares} {et~al.}(2025){Soares}, {Adibekyan}, {Mordasini}, {Deal}, {Sousa}, {Delgado-Mena}, {Santos}, \& {Dorn}}]{Soares2025}
{Soares}, B.~M.~T.~B., {Adibekyan}, V., {Mordasini}, C., {et~al.} 2025, \bibinfo{title}{{Assessing the processes behind planet engulfment and its imprints},} \aap, 693, A47, \dodoi{10.1051/0004-6361/202451399}

\bibitem[{M. {Soares-Furtado} {et~al.}(2021){Soares-Furtado}, {Cantiello}, {MacLeod}, \& {Ness}}]{SoaresFurtado2021}
{Soares-Furtado}, M., {Cantiello}, M., {MacLeod}, M., \& {Ness}, M.~K. 2021, \bibinfo{title}{{Lithium Enrichment Signatures of Planetary Engulfment Events in Evolved Stars},} \aj, 162, 273, \dodoi{10.3847/1538-3881/ac273c}

\bibitem[{N. {Soker}(2020){Soker}}]{Soker2020}
{Soker}, N. 2020, \bibinfo{title}{{Efficiently Jet-powered Radiation in Intermediate-luminosity Optical Transients},} \apj, 893, 20, \dodoi{10.3847/1538-4357/ab7dbb}

\bibitem[{N. {Soker}(2023){Soker}}]{Soker2023}
{Soker}, N. 2023, \bibinfo{title}{{On the nature of the planet-powered transient event ZTF SLRN-2020},} \mnras, 524, L94, \dodoi{10.1093/mnrasl/slad086}

\bibitem[{N. {Soker} \& N. {Kaplan}(2021){Soker} \& {Kaplan}}]{Soker2021a}
{Soker}, N., \& {Kaplan}, N. 2021, \bibinfo{title}{{Explaining recently studied intermediate luminosity optical transients (ILOTs) with jet powering},} Research in Astronomy and Astrophysics, 21, 090, \dodoi{10.1088/1674-4527/21/4/90}

\bibitem[{N. {Soker} \& R. {Tylenda}(2006){Soker} \& {Tylenda}}]{Soker2006}
{Soker}, N., \& {Tylenda}, R. 2006, \bibinfo{title}{{Violent stellar merger model for transient events},} \mnras, 373, 733, \dodoi{10.1111/j.1365-2966.2006.11056.x}

\bibitem[{N.~H. {Soliman} \& P.~F. {Hopkins}(2025){Soliman} \& {Hopkins}}]{Soliman2025}
{Soliman}, N.~H., \& {Hopkins}, P.~F. 2025, \bibinfo{title}{{Are Stars Really Ingesting Their Planets? Examining an Alternative Explanation},} \apj, 979, 98, \dodoi{10.3847/1538-4357/ada1d5}

\bibitem[{L. {Spina} {et~al.}(2015){Spina}, {Palla}, {Randich}, {Sacco}, {Jeffries}, {Magrini}, {Franciosini}, {Meyer}, {Tautvai{\v{s}}ien{\.{e}}}, {Gilmore}, {Alfaro}, {Allende Prieto}, {Bensby}, {Bragaglia}, {Flaccomio}, {Koposov}, {Lanzafame}, {Costado}, {Hourihane}, {Lardo}, {Lewis}, {Monaco}, {Morbidelli}, {Sousa}, {Worley}, \& {Zaggia}}]{Spina2015}
{Spina}, L., {Palla}, F., {Randich}, S., {et~al.} 2015, \bibinfo{title}{{The Gaia-ESO Survey: chemical signatures of rocky accretion in a young solar-type star},} \aap, 582, L6, \dodoi{10.1051/0004-6361/201526896}

\bibitem[{J.~E. {Staff} {et~al.}(2016){Staff}, {De Marco}, {Wood}, {Galaviz}, \& {Passy}}]{Staff2016}
{Staff}, J.~E., {De Marco}, O., {Wood}, P., {Galaviz}, P., \& {Passy}, J.-C. 2016, \bibinfo{title}{{Hydrodynamic simulations of the interaction between giant stars and planets},} \mnras, 458, 832, \dodoi{10.1093/mnras/stw331}

\bibitem[{A.~P. {Stephan} {et~al.}(2020){Stephan}, {Naoz}, {Gaudi}, \& {Salas}}]{Stephan2020}
{Stephan}, A.~P., {Naoz}, S., {Gaudi}, B.~S., \& {Salas}, J.~M. 2020, \bibinfo{title}{{Eating Planets for Lunch and Dinner: Signatures of Planet Consumption by Evolving Stars},} \apj, 889, 45, \dodoi{10.3847/1538-4357/ab5b00}

\bibitem[{Q. {Sun} {et~al.}(2025){Sun}, {Ting}, {Liu}, {Wang}, {Anthony-Twarog}, {Twarog}, {Yang}, {Chen}, {Karakas}, {Xie}, \& {Yong}}]{Sun2025}
{Sun}, Q., {Ting}, Y.-S., {Liu}, F., {et~al.} 2025, \bibinfo{title}{{C3PO. III. On the Lithium Signatures following Planet Engulfment by Stars},} \apj, 978, 107, \dodoi{10.3847/1538-4357/ad8dc3}

\bibitem[{R.~A. {Tejada Arevalo} {et~al.}(2021){Tejada Arevalo}, {Winn}, \& {Anderson}}]{TejadaArevalo2021}
{Tejada Arevalo}, R.~A., {Winn}, J.~N., \& {Anderson}, K.~R. 2021, \bibinfo{title}{{Further Evidence for Tidal Spin-up of Hot Jupiter Host Stars},} \apj, 919, 138, \dodoi{10.3847/1538-4357/ac1429}

\bibitem[{D.~P. {Thorngren} {et~al.}(2021){Thorngren}, {Fortney}, {Lopez}, {Berger}, \& {Huber}}]{Thorngren2021}
{Thorngren}, D.~P., {Fortney}, J.~J., {Lopez}, E.~D., {Berger}, T.~A., \& {Huber}, D. 2021, \bibinfo{title}{{Slow Cooling and Fast Reinflation for Hot Jupiters},} \apjl, 909, L16, \dodoi{10.3847/2041-8213/abe86d}

\bibitem[{D.~P. {Thorngren} {et~al.}(2023){Thorngren}, {Lee}, \& {Lopez}}]{Thorngren2023}
{Thorngren}, D.~P., {Lee}, E.~J., \& {Lopez}, E.~D. 2023, \bibinfo{title}{{Removal of Hot Saturns in Mass-Radius Plane by Runaway Mass Loss},} \apjl, 945, L36, \dodoi{10.3847/2041-8213/acbd35}

\bibitem[{F.~X. {Timmes} \& F.~D. {Swesty}(2000){Timmes} \& {Swesty}}]{Timmes2000}
{Timmes}, F.~X., \& {Swesty}, F.~D. 2000, \bibinfo{title}{{The Accuracy, Consistency, and Speed of an Electron-Positron Equation of State Based on Table Interpolation of the Helmholtz Free Energy},} \apjs, 126, 501, \dodoi{10.1086/313304}

\bibitem[{E. {Tognelli} {et~al.}(2016){Tognelli}, {Prada Moroni}, \& {Degl'Innocenti}}]{Tognelli2016}
{Tognelli}, E., {Prada Moroni}, P.~G., \& {Degl'Innocenti}, S. 2016, \bibinfo{title}{{Effect of planet ingestion on low-mass stars evolution: the case of 2MASS J08095427-4721419 star in the Gamma Velorum cluster},} \mnras, 460, 3888, \dodoi{10.1093/mnras/stw1268}

\bibitem[{R. Townsend(2024)Townsend}]{Townsend2024}
Townsend, R. 2024, \bibinfo{title}{MESA SDK for Mac OS,}, 24.7.1 Zenodo, \dodoi{10.5281/zenodo.13768941}

\bibitem[{J.~D. {Turner} {et~al.}(2021){Turner}, {Ridden-Harper}, \& {Jayawardhana}}]{Turner2021}
{Turner}, J.~D., {Ridden-Harper}, A., \& {Jayawardhana}, R. 2021, \bibinfo{title}{{Decaying Orbit of the Hot Jupiter WASP-12b: Confirmation with TESS Observations},} \aj, 161, 72, \dodoi{10.3847/1538-3881/abd178}

\bibitem[{R. {Tylenda}(2005){Tylenda}}]{Tylenda2005a}
{Tylenda}, R. 2005, \bibinfo{title}{{Evolution of V838 Monocerotis during and after the 2002 eruption},} \aap, 436, 1009, \dodoi{10.1051/0004-6361:20052800}

\bibitem[{R. {Tylenda} {et~al.}(2005){Tylenda}, {Crause}, {G{\'o}rny}, \& {Schmidt}}]{Tylenda2005}
{Tylenda}, R., {Crause}, L.~A., {G{\'o}rny}, S.~K., \& {Schmidt}, M.~R. 2005, \bibinfo{title}{{V4332 Sagittarii revisited},} \aap, 439, 651, \dodoi{10.1051/0004-6361:20041581}

\bibitem[{R. {Tylenda} {et~al.}(2011){Tylenda}, {Hajduk}, {Kami{\'n}ski}, {Udalski}, {Soszy{\'n}ski}, {Szyma{\'n}ski}, {Kubiak}, {Pietrzy{\'n}ski}, {Poleski}, {Wyrzykowski}, \& {Ulaczyk}}]{Tylenda2011}
{Tylenda}, R., {Hajduk}, M., {Kami{\'n}ski}, T., {et~al.} 2011, \bibinfo{title}{{V1309 Scorpii: merger of a contact binary},} \aap, 528, A114, \dodoi{10.1051/0004-6361/201016221}

\bibitem[{F. {Valsecchi} {et~al.}(2015){Valsecchi}, {Rappaport}, {Rasio}, {Marchant}, \& {Rogers}}]{Valsecchi2015}
{Valsecchi}, F., {Rappaport}, S., {Rasio}, F.~A., {Marchant}, P., \& {Rogers}, L.~A. 2015, \bibinfo{title}{{Tidally-driven Roche-lobe Overflow of Hot Jupiters with MESA},} \apj, 813, 101, \dodoi{10.1088/0004-637X/813/2/101}

\bibitem[{F. {Valsecchi} {et~al.}(2014){Valsecchi}, {Rasio}, \& {Steffen}}]{Valsecchi2014}
{Valsecchi}, F., {Rasio}, F.~A., \& {Steffen}, J.~H. 2014, \bibinfo{title}{{From Hot Jupiters to Super-Earths via Roche Lobe Overflow},} \apjl, 793, L3, \dodoi{10.1088/2041-8205/793/1/L3}

\bibitem[{D. {Veras}(2016){Veras}}]{Veras2016}
{Veras}, D. 2016, \bibinfo{title}{{Post-main-sequence planetary system evolution},} Royal Society Open Science, 3, 150571, \dodoi{10.1098/rsos.150571}

\bibitem[{E. {Villaver} \& M. {Livio}(2007){Villaver} \& {Livio}}]{Villaver2007}
{Villaver}, E., \& {Livio}, M. 2007, \bibinfo{title}{{Can Planets Survive Stellar Evolution?},} \apj, 661, 1192, \dodoi{10.1086/516746}

\bibitem[{E. {Villaver} \& M. {Livio}(2009){Villaver} \& {Livio}}]{Villaver2009}
{Villaver}, E., \& {Livio}, M. 2009, \bibinfo{title}{{The Orbital Evolution of Gas Giant Planets Around Giant Stars},} \apjl, 705, L81, \dodoi{10.1088/0004-637X/705/1/L81}

\bibitem[{E. {Villaver} {et~al.}(2014){Villaver}, {Livio}, {Mustill}, \& {Siess}}]{Villaver2014}
{Villaver}, E., {Livio}, M., {Mustill}, A.~J., \& {Siess}, L. 2014, \bibinfo{title}{{Hot Jupiters and Cool Stars},} \apj, 794, 3, \dodoi{10.1088/0004-637X/794/1/3}

\bibitem[{P. Virtanen {et~al.}(2020)Virtanen, Gommers, Oliphant, Haberland, Reddy, Cournapeau, Burovski, Peterson, Weckesser, Bright, {van der Walt}, Brett, Wilson, Millman, Mayorov, Nelson, Jones, Kern, Larson, Carey, Polat, Feng, Moore, {VanderPlas}, Laxalde, Perktold, Cimrman, Henriksen, Quintero, Harris, Archibald, Ribeiro, Pedregosa, {van Mulbregt}, \& {SciPy 1.0 Contributors}}]{Virtanen2020}
Virtanen, P., Gommers, R., Oliphant, T.~E., {et~al.} 2020, \bibinfo{title}{{{SciPy} 1.0: Fundamental Algorithms for Scientific Computing in Python},} Nature Methods, 17, 261, \dodoi{10.1038/s41592-019-0686-2}

\bibitem[{S. {Vissapragada} {et~al.}(2022){Vissapragada}, {Chontos}, {Greklek-McKeon}, {Knutson}, {Dai}, {P{\'e}rez Gonz{\'a}lez}, {Grunblatt}, {Huber}, \& {Saunders}}]{Vissapragada2022}
{Vissapragada}, S., {Chontos}, A., {Greklek-McKeon}, M., {et~al.} 2022, \bibinfo{title}{{The Possible Tidal Demise of Kepler's First Planetary System},} \apjl, 941, L31, \dodoi{10.3847/2041-8213/aca47e}

\bibitem[{N.~N. {Weinberg} {et~al.}(2024){Weinberg}, {Davachi}, {Essick}, {Yu}, {Arras}, \& {Belland}}]{Weinberg2024}
{Weinberg}, N.~N., {Davachi}, N., {Essick}, R., {et~al.} 2024, \bibinfo{title}{{Orbital Decay of Hot Jupiters due to Weakly Nonlinear Tidal Dissipation},} \apj, 960, 50, \dodoi{10.3847/1538-4357/ad05c9}

\bibitem[{D. {Xie} {et~al.}(2023){Xie}, {Zhu}, {Guo}, {Liu}, \& {L{\"u}}}]{Xie2023}
{Xie}, D., {Zhu}, C., {Guo}, S., {Liu}, H., \& {L{\"u}}, G. 2023, \bibinfo{title}{{An alternative formation scenario for uranium-rich giants: engulfing an Earth-like planet},} \mnras, 524, 3705, \dodoi{10.1093/mnras/stad2097}

\bibitem[{R. {Yamazaki} {et~al.}(2017){Yamazaki}, {Hayasaki}, \& {Loeb}}]{Yamazaki2017}
{Yamazaki}, R., {Hayasaki}, K., \& {Loeb}, A. 2017, \bibinfo{title}{{Optical-infrared flares and radio afterglows by Jovian planets inspiraling into their host stars},} \mnras, 466, 1421, \dodoi{10.1093/mnras/stw3207}

\bibitem[{H.-L. {Yan} {et~al.}(2018){Yan}, {Shi}, {Zhou}, {Chen}, {Li}, {Zhang}, {Bi}, {Wu}, {Li}, {Guo}, {Liu}, {Gao}, {Zhang}, {Zhou}, {Li}, \& {Zhao}}]{Yan2018}
{Yan}, H.-L., {Shi}, J.-R., {Zhou}, Y.-T., {et~al.} 2018, \bibinfo{title}{{The nature of the lithium enrichment in the most Li-rich giant star},} Nature Astronomy, 2, 790, \dodoi{10.1038/s41550-018-0544-7}

\bibitem[{J. {Yana Galarza} {et~al.}(2021){Yana Galarza}, {L{\'o}pez-Valdivia}, {Mel{\'e}ndez}, \& {Lorenzo-Oliveira}}]{YanaGalarza2021}
{Yana Galarza}, J., {L{\'o}pez-Valdivia}, R., {Mel{\'e}ndez}, J., \& {Lorenzo-Oliveira}, D. 2021, \bibinfo{title}{{Evidence of Rocky Planet Engulfment in the Wide Binary System HIP 71726/HIP 71737},} \apj, 922, 129, \dodoi{10.3847/1538-4357/ac2362}

\bibitem[{J. {Yana Galarza} {et~al.}(2024){Yana Galarza}, {Reggiani}, {Ferreira}, {Lorenzo-Oliveira}, {Simon}, {McWilliam}, {Schlaufman}, {Miquelarena}, {Flores Trivigno}, \& {Jaque Arancibia}}]{YanaGalarza2024}
{Yana Galarza}, J., {Reggiani}, H., {Ferreira}, T., {et~al.} 2024, \bibinfo{title}{{Detailed Abundances of the Planet-hosting TOI-1173 A/B System: Possible Evidence of Planet Engulfment in a Very Wide Binary},} \apj, 974, 122, \dodoi{10.3847/1538-4357/ad697f}

\bibitem[{R. {Yarza} {et~al.}(2023){Yarza}, {Razo-L{\'o}pez}, {Murguia-Berthier}, {Everson}, {Antoni}, {MacLeod}, {Soares-Furtado}, {Lee}, \& {Ramirez-Ruiz}}]{Yarza2023}
{Yarza}, R., {Razo-L{\'o}pez}, N.~B., {Murguia-Berthier}, A., {et~al.} 2023, \bibinfo{title}{{Hydrodynamics and Survivability during Post-main-sequence Planetary Engulfment},} \apj, 954, 176, \dodoi{10.3847/1538-4357/acbdfc}

\bibitem[{S.~W. {Yee} {et~al.}(2020){Yee}, {Winn}, {Knutson}, {Patra}, {Vissapragada}, {Zhang}, {Holman}, {Shporer}, \& {Wright}}]{Yee2020}
{Yee}, S.~W., {Winn}, J.~N., {Knutson}, H.~A., {et~al.} 2020, \bibinfo{title}{{The Orbit of WASP-12b Is Decaying},} \apjl, 888, L5, \dodoi{10.3847/2041-8213/ab5c16}

\bibitem[{M. {Zhang} \& K. {Penev}(2014){Zhang} \& {Penev}}]{Zhang2014}
{Zhang}, M., \& {Penev}, K. 2014, \bibinfo{title}{{Stars Get Dizzy After Lunch},} \apj, 787, 131, \dodoi{10.1088/0004-637X/787/2/131}

\bibitem[{W. {Zhu} \& S. {Dong}(2021){Zhu} \& {Dong}}]{Zhu2021}
{Zhu}, W., \& {Dong}, S. 2021, \bibinfo{title}{{Exoplanet Statistics and Theoretical Implications},} \araa, 59, 291, \dodoi{10.1146/annurev-astro-112420-020055}

\end{thebibliography}

\end{document}